\PassOptionsToPackage{dvipsnames,table}{xcolor}
\documentclass[conference,compsoc]{IEEEtran}
\usepackage{xcolor}
\usepackage[T1]{fontenc}
\usepackage[utf8]{inputenc}
\ifCLASSOPTIONcompsoc
  \usepackage[nocompress]{cite}
\else
  \usepackage{cite}
\fi
\usepackage{url}

\usepackage{xspace}
\usepackage{graphicx}
\usepackage{listings}
\usepackage{multirow}
\usepackage{pgfplots,pgfplotstable}
\usepackage{pifont}
\usepackage{pdfpages}
\usepackage{booktabs}
\usepackage[colorinlistoftodos]{todonotes}
\usepackage[normalem]{ulem}
\usepackage{zlmtt}
\usepackage[fixed]{fontawesome5}
\usepackage[flushleft]{threeparttable}
\usepackage{colortbl}
\usepackage{tikz}
\usetikzlibrary{tikzmark,shapes,calendar,matrix,calc,fit,decorations,
  overlay-beamer-styles,positioning,snakes,decorations.pathreplacing,
  backgrounds,patterns,arrows.meta,shadows}
\usepackage{pgf-pie}
\usepackage{forest}
\usepackage{adjustbox}
\usepackage{enumitem}
\usepackage[htt]{hyphenat}
\usepackage{amsmath,amssymb}

\usepackage{amsthm}
\theoremstyle{remark}
\newtheorem*{remark}{Remark}

\usepackage{hyperref}
\hypersetup{colorlinks, linkcolor={green!80!black}, citecolor={red!70!black}, urlcolor={blue!70!black}}

\newcommand{\ToolName}{Granite}
\newcommand{\etal}{et al. }

\newcommand{\tightpar}[1]{{\vspace{2pt}\noindent\bf #1. }}
\newcommand{\ie}[0]{i.e., }
\newcommand{\eg}[0]{e.g., }
\newcommand{\padtext}[2]{{\hspace*{#2}{#1}\hspace*{#2}}}

\def\printonlylargeenough#1#2{\unless\ifdim#2pt<#1pt\relax
	#2\printnumbertrue
	\else
	\printnumberfalse
	\fi}
\newif\ifprintnumber

\tikzset{piechart/.style={draw=none},}

\definecolor{OverprivilegedJobsColor}{HTML}{C94500}
\definecolor{MulitStepJobsColor}{HTML}{FFC154}
\definecolor{SingleStepJobsColor}{HTML}{98A737}
\definecolor{IgnoredJobsColor}{HTML}{868686}
\definecolor{AwesomeRed}{HTML}{ff2052}
\definecolor{MediumSeaGreen}{rgb}{0.24, 0.70, 0.44}
\definecolor{pullRequestsColor}{RGB}{228,26,28}  
\definecolor{contentsColor}{RGB}{55,126,184} 
\definecolor{issuesColor}{RGB}{77,175,74}   
\definecolor{actionsColor}{RGB}{152,78,163} 
\definecolor{securityEventsColor}{RGB}{255,127,0} 
\definecolor{checksColor}{RGB}{166,86,40} 

\definecolor{dgray} {RGB}{220,220,220}   %
\definecolor{dramp0}{RGB}{255,243,220}   %
\definecolor{dramp1}{RGB}{255,224,178}   %
\definecolor{dramp2}{RGB}{255,193,120}   %
\definecolor{dramp3}{RGB}{249,156, 90}   %
\definecolor{dramp4}{RGB}{240,120, 80}   %
\definecolor{dramp5}{RGB}{225, 90, 70}   %
 
\definecolor{dtext} {RGB}{ 90, 40, 10}   %
\definecolor{dgtext}{RGB}{ 80, 80, 80}   %

\definecolor{exCap}{RGB}{204,85,0}
\definecolor{exComment}{RGB}{110,110,110}
\definecolor{exKeyword}{RGB}{0,92,170}
\definecolor{exString}{RGB}{0,128,64}
\tikzset{exstepnode/.style={circle, draw=black, line width=0.35pt, fill=white, inner sep=0pt, minimum size=3.2mm, font=\bfseries\fontsize{6pt}{6pt}\selectfont}}
\newcommand{\exstep}[1]{\tikz[baseline={([yshift=-0.3ex]n.base)}]\node[exstepnode] (n) {#1};}
\usepackage{xcolor}
\usepackage{listings}
\usepackage{tikz}
\usetikzlibrary{calc}
\colorlet{punct}{red!60!black}
\definecolor{background}{HTML}{EEEEEE}
\definecolor{delim}{RGB}{20,105,176}
\colorlet{numb}{magenta!60!black}
\definecolor{commentgreen}{RGB}{176,176,176}
\definecolor{rowcolor}{cmyk}{0,0.87,0.68,0.32}
\definecolor{rowcolor2}{cmyk}{0.20,0,0.37,0.34}
\definecolor{eminence}{RGB}{108,48,130}
\definecolor{weborange}{RGB}{255,165,0}
\definecolor{frenchplum}{RGB}{129,20,82}
\definecolor{darkgreen}{RGB}{10,92,10}
\definecolor{celadon}{RGB}{172,225,175}
\definecolor{jsonString}{HTML}{1560BD}
\definecolor{jsonKeyword}{HTML}{fc1e70}

\definecolor{ghblue}{RGB}{0,92,197}           %
\definecolor{ghstring}{RGB}{3,47,98}          %
\definecolor{ghgreen}{RGB}{34,134,58}         %
\definecolor{ghgray}{RGB}{106,115,125}        %
\definecolor{ghnumber}{RGB}{5,80,174}         %
\definecolor{ghbg}{RGB}{246,248,250}          %

\makeatletter
\newenvironment{btHighlight}[1][]
  {\begingroup\tikzset{bt@Highlight@par/.style={#1}}\begin{lrbox}{\@tempboxa}}
  {\end{lrbox}\bt@HL@box[bt@Highlight@par]{\@tempboxa}\endgroup}
\newcommand\btHL[1][]{%
  \begin{btHighlight}[#1]\bgroup\aftergroup\bt@HL@endenv%
}
\def\bt@HL@endenv{%
  \end{btHighlight}%
  \egroup
}
\newcommand{\bt@HL@box}[2][]{%
  \tikz[#1]{%
    \pgfpathrectangle{\pgfpoint{1pt}{0pt}}{\pgfpoint{\wd #2}{\ht #2}}%
    \pgfusepath{use as bounding box}%
    \node[anchor=base west, fill=orange!30, outer sep=0pt, inner xsep=1pt, 
          inner ysep=0pt, rounded corners=3pt, minimum height=\ht\strutbox+1pt, #1]
          {\raisebox{1pt}{\strut}\strut\usebox{#2}};
  }%
}
\makeatother
\lstdefinelanguage{YAML}{
    basicstyle=\fontsize{7}{7}\ttfamily,
    numberstyle=\scriptsize\color{ghgray},
    tabsize=2,
    numbers=left,
    stepnumber=1,
    numbersep=8pt,
    showstringspaces=false,
    breaklines=true,
    xleftmargin=12pt,
    alsoletter={-},
    morekeywords=[1]{true,false,null,yes,no,off},
    morekeywords=[2]{name, runs, on, runs-on, run, jobs, steps, uses, env, needs, with, if, id, strategy, matrix, services, outputs, secrets, permissions, branches, paths, workflow_dispatch, push, pull_request, pull-requests, schedule, types, checks, contents, issues, packages, deployments, id-token, timeout-minutes, continue-on-error, working-directory, shell, defaults, concurrency, cancel-in-progress},
    keywordstyle=[1]\color{ghblue}\bfseries,
    keywordstyle=[2]\color{ghblue}\bfseries,
    identifierstyle=\color{black},
    sensitive=true,
    comment=[l]{\#},
    commentstyle=\color{ghgray}\itshape,
    morestring=[b]',
    morestring=[b]",
    stringstyle=\color{ghstring},
    literate=
     *{0}{{{\color{black}0}}}{1}
      {1}{{{\color{black}1}}}{1}
      {2}{{{\color{black}2}}}{1}
      {3}{{{\color{black}3}}}{1}
      {4}{{{\color{black}4}}}{1}
      {5}{{{\color{black}5}}}{1}
      {6}{{{\color{black}6}}}{1}
      {7}{{{\color{black}7}}}{1}
      {8}{{{\color{black}8}}}{1}
      {9}{{{\color{black}9}}}{1}
      {:}{{{\color{black}{:}}}}{1}
      {,}{{{\color{black}{,}}}}{1}
      {?}{{{\color{ghgreen}{?}}}}{1}
      {!}{{{\color{ghgreen}{!}}}}{1}
      {\{}{{{\color{ghblue}{\{}}}}{1}
      {\}}{{{\color{ghblue}{\}}}}}{1}
      {[}{{{\color{ghblue}{[}}}}{1}
      {]}{{{\color{ghblue}{]}}}}{1}
      {|}{{{\color{ghgreen}{|}}}}{1}
      {>}{{{\color{ghgreen}{>}}}}{1}
      {@}{{{\color{black}{@}}}}{1}
      {\ on:}{{{\color{ghblue}\bfseries\ on:}}}{4}
      {\ on\ }{{{\color{ghblue}\bfseries\ on\ }}}{4},
    moredelim=[is][\color{ghgreen}]{LISTHYPHEN}{LISTHYPHEN},
    moredelim=**[is][{\btHL[fill=weborange!40]}]{~}{~},
    moredelim=**[is][{\btHL[fill=celadon!40]}]{!}{!},
    moredelim=**[is][{\btHL[fill=frenchplum!40]}]{±}{±},
    keepspaces=true
}
\lstdefinelanguage{JavaScript}{
    basicstyle=\fontsize{6.7}{7}\ttfamily,
    numberstyle=\scriptsize\color{gray},
    tabsize=2,
    numbers=left,
    stepnumber=1,
    numbersep=8pt,
    xleftmargin=12pt,
    showstringspaces=false,
    breaklines=true,
    keywords={typeof,new,true,false,catch,function,return,null,switch,var,let,const,if,in,while,do,else,case,break},
    keywordstyle=\color{blue}\bfseries,
    ndkeywords={class,export,boolean,throw,implements,import,this},
    ndkeywordstyle=\color{eminence}\bfseries,
    identifierstyle=\color{black},
    sensitive=true,
    comment=[l]{//},
    morecomment=[s]{/*}{*/},
    commentstyle=\color{commentgreen}\ttfamily,
    morestring=[b]',
    morestring=[b]",
    stringstyle=\color{frenchplum}\ttfamily,
    keepspaces=true
}
\lstdefinelanguage{JSON}{
	basicstyle=\fontsize{6.5}{7}\ttfamily,
    numberstyle=\scriptsize\color{gray},
    tabsize=2,
    numbers=left,
    stepnumber=1,
    numbersep=8pt,
    xleftmargin=12pt,
    showstringspaces=false,
    breaklines=true,
    string=[s]{"}{"},
    stringstyle=\color{jsonString},
    keywords={true,false,null},
    keywordstyle=\color{jsonKeyword}\bfseries,
    literate=
     *{0}{{{\color{numb}0}}}{1}
      {1}{{{\color{numb}1}}}{1}
      {2}{{{\color{numb}2}}}{1}
      {3}{{{\color{numb}3}}}{1}
      {4}{{{\color{numb}4}}}{1}
      {5}{{{\color{numb}5}}}{1}
      {6}{{{\color{numb}6}}}{1}
      {7}{{{\color{numb}7}}}{1}
      {8}{{{\color{numb}8}}}{1}
      {9}{{{\color{numb}9}}}{1}
      {:}{{{\color{punct}{:}}}}{1}
      {,}{{{\color{punct}{,}}}}{1}
      {\{}{{{\color{delim}{\{}}}}{1}
      {\}}{{{\color{delim}{\}}}}}{1}
      {[}{{{\color{delim}{[}}}}{1}
      {]}{{{\color{delim}{]}}}}{1},
    keepspaces=true
}
\lstdefinestyle{yaml}{
    language=YAML,
    breaklines=true
}

\let\origunderscore\_
\renewcommand{\_}{\origunderscore\discretionary{}{}{}}

\newif\iffull
\fulltrue

\begin{document}

\title{Granite: Granular Runtime Enforcement for GitHub Actions Permissions}
\author{%
  \IEEEauthorblockN{Mojtaba Moazen\IEEEauthorrefmark{1}, Amir M. Ahmadian\IEEEauthorrefmark{2}, Musard Balliu\IEEEauthorrefmark{1}}%
  \IEEEauthorblockA{\IEEEauthorrefmark{1}KTH Royal Institute of Technology, Sweden\\
  \IEEEauthorrefmark{2}Chalmers University of Technology and University of Gothenburg, Sweden\\
  \{moazen, musard\}@kth.se, amir.ahmadian@chalmers.se}%
}
\maketitle

\begin{abstract}
Modern software projects use automated CI/CD pipelines and GitHub Actions is a popular platform for building them.  It enables project maintainers to create custom workflows that compose reusable third-party actions into jobs of sequential steps. Wary of the security risks of fully-privileged actions, GitHub enforces job-level token permissions for controlling workflow access to the repository resources. This permission model violates the principle of least privilege and can lead to broader software supply chain attacks. Every step in a job shares the full permissions of that job, hence a single compromised action can misuse all permissions, as witnessed by the recent tj-actions compromise. 

In this paper, we present {\ToolName}, the first system to enforce fine-grained permissions for GitHub Actions at the step-level granularity at runtime. {\ToolName} ensures that untrusted steps never hold the real token. Instead, a separate controller issues a step identifier token and provides the real token only to the requests that meet the step-level policy. It builds on GitHub’s self-hosted runner and requires no changes to the workflows, while  handling JavaScript, composite, and Docker actions transparently.
We evaluate {\ToolName} for compatibility, security, and performance using a dataset of 500 workflows comprising 12,915 jobs from the most-starred GitHub repositories that use GitHub Actions. Our analysis reveals that 46.3\% of the jobs are overprivileged and can be protected by {\ToolName}.  We evaluate compatibility  on 46 repositories and 127 unique actions, finding that {\ToolName} preserves the behavior of every job that runs on the unmodified runner, except one that it blocks by design. 
We also validate the attack prevention capabilities of  {\ToolName} , including attempts to bypass its enforcement,  by reproducing 10 permission misuse attacks over 42 overprivileged jobs, and show that it removes write access from approximately 83.3\% of the steps that hold it under GitHub's current model. Our performance evaluation shows that {\ToolName} introduces an average per-step overhead of 7.9\%. These results indicate that {\ToolName} is a practical and effective tool to reduce the attack surface in CI/CD pipelines.
\end{abstract}
\IEEEpeerreviewmaketitle

\section{Introduction}\label{sec:introduction}

GitHub Actions is a popular CI/CD (Continuous Integration{\slash}Con\-tin\-u\-ous Delivery) automation platform built into GitHub that enables repository maintainers to run customizable workflows to build, test, and deploy code directly from the repository~\cite{GithubActions}. 
Workflows are structured files that define when and how automation runs by organizing \emph{tasks} into sequential or parallel jobs, where each job is made up of \emph{steps} that execute specific \emph{actions} such as testing or deploying the application (Listing \ref{listing:running_example} shows an example). 
Actions are reusable third-party components written in JavaScript, Bash or as a Docker container, allowing them to run commands or execute scripts across different operating systems. 
Recent works \cite{DecanICSME,PabloSenar,DECAN2023} show that the use of third-party actions in workflows is pervasive.  

Unsurprisingly, the convenience of reusability and automation comes with security risks. 
In this setting, the risk is further exacerbated by the central role of workflows in accessing and managing the entire pipeline with high privileges over sensitive resources such as the code repository. 
This makes them an appealing target for software supply chain attacks \cite{LadisaSP,SSCSurvey,Quantifying,argus}. 
In fact, Koishybayev \etal \cite{Characterizing} provide alarming evidence of the issue, with $99.8\%$ of workflows being overprivileged with read-write access (instead of read-only) on the repository. 
These overprivileges are being exploited by attackers \cite{CWE732Incorrect,CWE250Execution,CWE269Improper,nistCVE202349291}, as witnessed by the infamous supply chain attack on \texttt{tj-actions/changed-files} action \cite{githubCVE202530066}. More recently, compromised packages misused workflow permissions to leak stolen secrets through the victims' repositories~\cite{stepsecurityRedHat}.

Wary of the risks of overprivileged workflows, GitHub recently introduced a permission system aiming at enforcing the principle of least privilege~\cite{saltzer}, along with the GitHub Monitor and Advisor system \cite{GithubPer}, on a key resource: \emph{the code repository}. 
The system monitors the actual permissions used by a job in a workflow at runtime and reports them to the developer. 
Unfortunately, as we show in this paper, it fails to provide context-dependent fine-grained permissions at the level of steps, enabling any action within a job to use the permissions granted to that job. 
Other proposals \cite{stepsecurityKnowledgeBase,cosseter} suggest manual and static analysis to infer minimal set of permissions required by an action. 
Beyond the known challenges with analyzing (obfuscated) JavaScript and Bash code, these approaches suffer from the lack of contextual information, \eg trigger events and input parameters. Consequently, they may include permissions associated with paths not exercised by a particular workflow or omit permissions that they fail to resolve (Section~\ref{sec:evaluations}).

This paper aims to address these limitations by designing, implementing, and evaluating {\ToolName}, the first system to enforce fine-grained permissions for GitHub workflows. 
{\ToolName} builds on the GitHub's self-hosted runner as the execution environment and ensures that untrusted steps never hold the real token.
Instead, each step receives a step identifier token (SIT), and a separate controller provides the real token only to the GitHub requests that are in line with the step's fine-grained permissions.
{\ToolName} provides seamless integration with a GitHub-hosted repository. %

Our evaluation confirms the coarse-grained nature of the GitHub's current permission system by analyzing 500 most starred workflows (and about 13,000 jobs) from GitHub repositories, showing that overprivileged jobs are prevalent ($46.3\%$).
We then select 48 jobs using 127 distinct actions from 46 popular repositories to conduct a thorough evaluation of {\ToolName} with regard to compatibility and a comparison of observed permissions with state-of-the-art tools such as GitHub Advisor~\cite{GithubPer}, Step Security's permission knowledge base~\cite{stepsecurityKnowledgeBase} and COSSETER~\cite{cosseter}.
We further use {\ToolName} to identify 42 overprivileged jobs (with about $59.0\%$ of High/Critical severity) and implement proof-of-concept exploits to show their feasibility under the current permission model. 
We confirm that {\ToolName} prevents the attacks, including attempts to bypass its enforcement, and reduces the attack surface by approximately $83.3\%$ per overprivileged job.
Finally, our performance evaluation shows that {\ToolName} introduces an average overhead of $8.5\%$ for JavaScript actions, $6.4\%$ for composite actions, $6.6\%$ for run steps, and $1.9\%$ for Docker actions. 
These results show that {\ToolName} is an effective and efficient tool to enforce the principle of least privilege on GitHub workflows.

\noindent In summary, we offer the following contributions:
\begin{itemize}[leftmargin=*]
    \item We design a system that enforces context-dependent fine-grained policies for GitHub workflows at step-level granularity, redesigning GitHub's permission model to reduce the attack surface against code repository (Section~\ref{sec:tool_design}).
    \item We implement {\ToolName}, the first runtime permission system for GitHub Actions able to enforce the minimum permissions for each workflow step at runtime (Section~\ref{sec:implementation}).
    \item We systematically evaluate {\ToolName} on real-world workflows in terms of prevalence, compatibility, security, and performance~(Section~\ref{sec:evaluations}).
\end{itemize}

\section{A Primer on GitHub Actions}\label{sec:github_actions}
This section provides an introduction to the GitHub workflows, their permission model, and the GitHub runner.

\subsection{GitHub Workflows}
GitHub workflows are automated pipelines designed to streamline software development and CI/CD processes directly into a GitHub repository.
A \emph{workflow} is defined via a \texttt{YAML} file stored within the repository's root directory, specifying a series of \emph{jobs} that execute in response to specific \emph{trigger events}, such as code pushes, pull requests, or issue creation.
Each job is defined by a sequence of \emph{steps} that are executed sequentially, while jobs themselves can be executed concurrently or sequentially.

Listing \ref{listing:running_example} illustrates how GitHub workflows function on a modified workflow
from \texttt{activeadmin}~\cite{activeadmin}, a popular Ruby on Rails framework for creating admin dashboards. 
Upon a pull-request event, \texttt{activeadmin} runs Markdown linting checks using the \texttt{markdownlint} job (line~\ref{lst_line:job_mdlint}).

\tightpar{Jobs}%
A job and its steps are executed within a runner platform on an environment specified by the \texttt{runs-on} keyword.
For example, the job \texttt{markdownlint} executes on the latest Ubuntu version (line~\ref{lst_line:runs_on}).
The runner itself is either self-hosted or provided by GitHub.

\lstset{
	language=YAML,
	breaklines=true, 
	style=yaml,
	firstnumber=1,
	escapeinside={*|}{|*},
	postbreak=\mbox{\textcolor{red}{$\hookrightarrow$}\space},
}
\begin{lstlisting}[label=listing:running_example,caption={Markdown Lint workflow from \texttt{activeadmin}}, captionpos=b]{YAML}
name: Markdown Lint, Build, and Test
on: 													*|\label{lst_line:trigger_on}|*
	pull_request:
	push:
env:
	...
jobs:
	markdownlint: *|\label{lst_line:job_mdlint}|*
		~permissions~                					*|\label{lst_line:permissions}|*
			~contents: read~
			~pull-requests: write~
		name: Run markdownlint
		runs-on: ubuntu-latest 							*|\label{lst_line:runs_on}|*
		if: github.event_name == 'pull_request'	
		steps:
			- uses: actions/checkout@v5
			- uses: tj-actions/changed-files@v47
			...
			- uses: reviewdog/action-markdownlint@v0
			...
	build:
		name: Build
		runs-on: ubuntu-latest
		if: github.event_name == 'push'
		steps:
			- uses: actions/checkout@v5
			- uses: ruby/setup-ruby@v1 *|\label{lst_line:ruby_setup}|*
			  with:
					ruby-version: '3.4'
			- name: Run Ruby build composite action
			  uses:./.github/actions/ruby-build *|\label{lst_line:composite}|*
			- name: call reusable workflow
			  uses: ./.github/workflows/create_automerge.yml*|\label{lst_line:call_reusable_workflow}|*
			  with:
					....
	test:
		name: Test after build
		needs : build
		runs-on: ubuntu-latest
		steps:
			- uses: actions/checkout@v5
			- name: Run tests
			  run: bundle exec rake test 					*|\label{lst_line:run_direct}|*
			  shell: bash
\end{lstlisting}

Jobs do not share data by default but instead use artifacts, \eg environment variables, to exchange data. %
By default, jobs will run simultaneously, each within a separate instance of the runner.
When needed, jobs can be executed sequentially using the keyword \texttt{needs} in a \texttt{YAML} file.
For instance, the \texttt{test} job will only execute after the completion of the \texttt{build} job.
Jobs can also be conditioned on specific events, in Listing \ref{listing:running_example} job \texttt{build} only runs after a push event, while job \texttt{markdownlint} runs upon a pull request.

\tightpar{Steps}%
A step is an individual task executed within a job's execution context.
Jobs use separate steps to define and execute tasks such as checking out the code, installing dependencies, running tests, and deploying the software.
All steps in a job share the job's execution context, including secrets, and use the filesystem for passing data. 
Unlike jobs, steps are \emph{always} executed sequentially. %

In GitHub workflows, steps are executed as either commands or actions. 
Commands, specified with the \texttt{run} keyword, allow developers to execute shell-based instructions directly within the workflow. 
For example, in Listing \ref{listing:running_example}, on line \ref{lst_line:run_direct}, the runner executes the command \texttt{bundle} \texttt{exec} \texttt{rake} \texttt{test} on a bash shell.
Actions, on the other hand, are specified by the \texttt{uses} keyword and represent reusable components.
A JavaScript action, which is the most common type of action, executes custom code that is defined in an \texttt{action.yml} file pointing to the main JavaScript entrypoint.
A composite action bundles multiple workflow steps, such as running shell commands or invoking other actions, into a single reusable action.
A Docker action runs its code in a container built from a Dockerfile or a published image. 
These actions can be sourced from the GitHub Marketplace~\cite{GitHubMarketplace} or defined internally.

Job \texttt{build} in Listing \ref{listing:running_example} has four steps.
The \texttt{actions/checkout} action which checks out the repository content is an example of an action obtained from GitHub Marketplace, whereas \texttt{ruby-build} on line \ref{lst_line:composite} is an instance of a (composite) action  defined internally and is responsible for building the program from its Ruby source code.
Line \ref{lst_line:ruby_setup} shows a reusable action (\texttt{actions/setup-ruby}) written in JavaScript that installs Ruby version $3.4$ on the runner. 
Workflows defined on the \texttt{workflow\_call} event can be invoked as reusable workflows, as shown on line \ref{lst_line:call_reusable_workflow}.

\subsection{Runner Execution}
GitHub runs workflows on its hosted runners and also offers an open-source runner that can be self-hosted after registering it with a repository, organization, or project.
When a workflow is triggered, the GitHub Actions service schedules its jobs. It then sends each job's metadata (encrypted), including the steps to execute, to the runner's job listener.
For each job, the runner initiates a worker process to execute its steps. Shell commands and JavaScript actions run as subprocesses of the worker, Docker actions run in containers, and composite actions execute their nested steps in sequence.
The job's metadata also includes the \texttt{GITHUB\_TOKEN}, a temporary credential that the runner exposes to steps through the \texttt{github} and \texttt{secrets} contexts.
This sensitive token provides scoped permissions to the repository, enabling the job to perform authenticated operations such as cloning the repository or invoking GitHub's REST APIs, depending on the permissions configured. This token is invalidated once the job completes.

\tightpar{GitHub token permissions}%
The permissions of \texttt{GITHUB\allowbreak\_TOKEN} guard access to \emph{16} content types within the repository. 
Table~\ref{tab:github_permissions_risk}  shows their access levels and risks for public repositories, and thus a lower bound for private ones.

\begin{table*}[t]
	\centering
	\caption{GitHub Actions Permissions and Their Risk assessment}
	\label{tab:github_permissions_risk}
	\rowcolors{3}{gray!10}{}
	\scalebox{0.79}{%
	\begin{tabular}{l c c l c l}
		& & \multicolumn{2}{c}{\textbf{Read}} & \multicolumn{2}{c}{\textbf{Write}} \\
		\cmidrule(lr){3-4} \cmidrule(lr){5-6}
		& \textbf{Allowed Values} & \padtext{Risk}{20pt} & Effect & \padtext{Risk}{20pt} & Effect \\   
		\midrule[1px]
		\textbf{contents} & read \textbar{} write \textbar{} none & Low & View repository contents & Critical & Modify code, inject backdoor \\
		\textbf{deployments} & read \textbar{} write \textbar{} none & Low & View deployment history & Critical & Trigger or alter deployments \\
		\textbf{packages} & read \textbar{} write \textbar{} none & Low & Download packages & High & Publish, replace, or poison packages \\
		\textbf{pull-requests} & read \textbar{} write \textbar{} none & Low & View pull requests & High & Open or merge malicious pull requests \\
		\textbf{security-events} & read \textbar{} write \textbar{} none & Medium & View vulnerability data & High & Suppress alerts, hide vulnerabilities \\
		\textbf{actions} & read \textbar{} write \textbar{} none & Low & View workflows & High & Modify workflows, escalate privileges \\
		\textbf{checks} & read \textbar{} write \textbar{} none & Low & View CI check results & Medium & Fabricate CI outcomes \\
		\textbf{statuses} & read \textbar{} write \textbar{} none & Low & View commit/build statuses & Medium & Spoof commit/build statuses \\
		\textbf{issues} & read \textbar{} write \textbar{} none & Low & View discussions for intel & Low & Spam, social engineering \\
		\textbf{attestations} & read \textbar{} write \textbar{} none & Low & View attestation metadata & Medium & Tamper with supply chain attestations \\
		\textbf{discussions} & read \textbar{} write \textbar{} none & Low & View community discussions & Low & Post misleading or malicious content \\
		\textbf{pages} & read \textbar{} write \textbar{} none & Low & View published pages & Medium & Deploy defaced or malicious pages \\
		\textbf{artifact-metadata} & read \textbar{} write \textbar{} none & Low & View artifact storage/deployment records & Medium & Forge storage or deployment records \\
		\textbf{code-quality} & read \textbar{} write \textbar{} none & Low & View code quality findings & Low & Upload falsified coverage/quality reports \\
		\textbf{copilot-requests} & \phantom{read \textbar{} }write \textbar{} none & - & - & Medium & Abuse organization-billed Copilot quota \\
		\textbf{vulnerability-alerts} & read \textbar{} \phantom{write \textbar{} }none & Medium & View unpatched dependency vulnerabilities & - & - \\
	\end{tabular}}
\end{table*} 
		 
These permissions can be configured in the repository settings or declared in the workflow file, either for the entire workflow or for individual jobs, but not for individual steps.
When declared at the job level, the permissions must be specified at the beginning of the job definition. 
Each permission in Table~\ref{tab:github_permissions_risk} can be set to \texttt{read} (if applicable), \texttt{write} (which implies \texttt{read}), or \texttt{none}. 
Once any permission is declared, unspecified permissions default to \texttt{none}, otherwise, the repository's default applies, which grants \texttt{write} access to all scopes.
For example, the \texttt{markdownlint} job in Listing~\ref{listing:running_example} is granted only \texttt{contents: read} and \texttt{pull-requests: write} (line~\ref{lst_line:permissions}). The latter also implies read access, and all three steps share these permissions.
Workflows may instead use a personal access token (\texttt{PAT}) stored as a repository secret.  GitHub recommends \texttt{GITHUB\_TOKEN}, which is temporary and  repository-scoped, whereas a \texttt{PAT} is long-lived and covers multiple repositories, which increases the security risks~\cite{GithubTokenNote}.
Like existing works~\cite{GithubPer,stepsecurityKnowledgeBase,cosseter}, we focus on permission scopes of  \texttt{GITHUB\_TOKEN}. 

\section{Problem Statement and Threat Model}\label{sec:problem}
GitHub provides a centralized permission model to regulate access to the code repository. It enforces \texttt{GITHUB\_TOKEN} permissions
at the granularity of workflows or jobs, hence every step can use permissions that only its neighbors need. This design choice violates the least privilege principle, enabling a single compromised step to misuse all its job permissions. We illustrate the problem, discuss the goals of step-level enforcement, and define our threat model.

\subsection{Motivating Example}\label{Sec:Motivating_example}
We illustrate the problem of overprivileged actions via the workflow in Listing \ref{listing:running_example}.
The  \texttt{markdownlint} job has three steps. The \texttt{actions/checkout} action checks out the repository code and needs \emph{contents: read} permission on the content of the repository. \texttt{tj-actions/changed-files} determines which Markdown files have been modified by a pull request and needs \emph{pull-requests: read}. \texttt{reviewdog/action-markdownlint} lints these files and reports the findings on the pull request, which needs \emph{pull-requests: write}.
Following GitHub's best practice~\cite{githubPermissionSyntax}, the developer declares \emph{contents: read} and \emph{pull-requests: write} permissions for the job (line \ref{lst_line:permissions}). %
These permissions apply to all three steps, therefore \texttt{tj-actions/changed-files} also holds \emph{pull-requests: write}, which  it does not need. %
An attacker who compromises this action can use that permission, for example to approve pull requests on the repository (Listing \ref{listing:tjchangedfiles}),  post misleading reviews, or edit and close pull requests.

	\lstset{
	language=JavaScript,
	breaklines=true,
	firstnumber=1,
	basicstyle=\fontsize{6.5}{7}\ttfamily,
	postbreak=\mbox{\textcolor{red}{$\hookrightarrow$}\space},
}
\begin{lstlisting}[label=listing:tjchangedfiles,caption={Approve a pull-request by tj-changed-files actions}, captionpos=b]
await octokit.rest.pulls.createReview({
	owner,
	repo,
	pull_number,
	event: 'APPROVE',
	body: 'Approve the pull-request.',});
\end{lstlisting}
Such attacks on GitHub Actions are real. In March 2025, the same \texttt{tj-actions/changed-files} action was compromised. 
The attacker retroactively pointed multiple version tags to a malicious commit that dumped the memory of the runner’s worker process and printed sensitive credentials (\eg API keys, npm tokens, and private keys) to the workflow logs, affecting an estimated  $23,000$ repositories~\cite{sysdig,cveCVE202530066}.
In 2026, compromised npm packages executed by a workflow step similarly harvested the job’s  \texttt{GITHUB\_TOKEN}  and used its  \emph{contents: write} permission to commit the stolen secrets to the victims' repositories~\cite{stepsecurityRedHat}. In both cases, a single step exploited the privileges of its underlying  job. 

The exposure is larger when developers do not follow best practices as is often the case. For example, the original \texttt{activeadmin} workflow~\cite{activeadmin}  has no permissions block, hence the token carries the repository's default permissions, which can include write access to most scopes, and every step can use all of them.

\subsection{Step-level Permissions}
\label{sec:problem_setting}
The risk of job-level permission is high because steps typically use actions developed by third parties~\cite{githubSecureReference}. An immediate solution is to put each step in a separate job. However, unlike steps, jobs do not share their execution context and must pass data explicitly, which complicates the design and maintenance of workflows. Another solution is to give each step its own narrow-scoped token, \eg a separate \texttt{PAT} passed only to the step that needs it. Beyond the burden of creating new token per step, this does not help either, because the runner still exposes the job's \texttt{GITHUB\_TOKEN} to every step through the \texttt{github} context, \eg as the default value of an action's input, even when the workflow never passes it.

We therefore enforce step-level permissions in the runner itself, by keeping the  \texttt{GITHUB\_TOKEN} away from steps and validating their requests to GitHub at runtime. Since GitHub's hosted environment is not open to modification, we build on the self-hosted runner, which is open source. Such a system should account for the following goals and existing workflows should run unmodified.

\tightpar{Context-dependent permissions} 
The required permissions of a step  depend on the underlying workflow, not only on the action it invokes. Consider \texttt{tj-actions/changed-files}  from our motivating example (Listing~\ref{listing:tjIfcondition}).
	
\lstset{
	language=JavaScript,
	breaklines=true,
	firstnumber=1,
    escapeinside={*|}{|*},
    basicstyle=\fontsize{6.5}{7}\ttfamily,
	postbreak=\mbox{\textcolor{red}{$\hookrightarrow$}\space},
}
\begin{lstlisting}[label=listing:tjIfcondition,caption={ tj-actions/changed-files GitHub action }, captionpos=b]
if (!github.context.payload.pull_request?.base?.ref) { *|\label{tj_lst_line:if_condition}|*
	core.info('Run on non-PR event')
	diffResult = await getSHAForNonPullRequestEvent({
		....})
} else {
	core.info('Run on PR event')
	diffResult = await getSHAForPullRequestEvent({ *|\label{tj_lst_line:getSHAForPullRequestEvent}|*
	....})}
\end{lstlisting}

The action determines the files changed by the triggering event. For a pull-request event, \texttt{getSHAForPullRequestEvent} (line \ref{tj_lst_line:getSHAForPullRequestEvent}) calls GitHub API endpoints that need \emph{pull-requests: read}. For any other event,   \texttt{getSHAForNonPullRequestEvent} uses  \texttt{git diff}  and needs no permission. If the  \texttt{markdownlint} job in Listing~\ref{listing:running_example} also ran on \texttt{push}, its second step would thus need no permission in that context, yet both a fixed step-level policy and an action-level static analysis would grant it \emph{pull-requests: read}. Since the triggers, like the inputs passed to an action, are fixed in the workflow file, permissions must be determined per step and per triggering event of a workflow, rather than per action.

\tightpar{Coverage of all step types} 
 A step can reach GitHub in many ways. JavaScript actions use a variety of HTTP modules and libraries,  \texttt{run} steps and composite actions call command-line tools such as  \texttt{curl} or \texttt{wget}, and Docker actions execute their own programs. Moreover, not all requests go through the REST API. For example, pushing a commit uses Git's HTTP protocol, while pushing an image uses the container registry. Enforcement must therefore cover every request to GitHub, independently of the step type and the program that issues it.

\subsection{Threat Model}\label{Sub:threat_model}
Our threat model considers an attacker that controls the code of one or more steps of a job. This can be a third-party action, compromised through its supply chain~\cite{CIServices} or altered by its maintainer, or a dependency executed by a \texttt{run}  step~\cite{stepsecurityRedHat}. The attacker's goal is to use the job’s \texttt{GITHUB\_TOKEN} beyond the permissions granted to the step, breaking the integrity or confidentiality of the repository. Attacker code runs as the user of the runner's worker process, possibly as root inside the runner's container. It can read and write the job’s filesystem and environment, start processes, and open arbitrary network connections. Besides calling the GitHub API directly, the attacker may try to read the token from the runner, impersonate the runner towards \ToolName, reuse a credential issued to another step, or contact GitHub while bypassing \ToolName. Section \ref{sec:tool_design} states the design goals against each attempt, and Section \ref{Sub:RQ3} tests them.

We trust GitHub, the \ToolName{}   runner and controller, the top-level workflow file, and the step-level policy, which developers should review before enforcing it (Section \ref{sec:limitations}). We assume that steps cannot escape the runner's container or modify the runner's own files, and that a runner executes a single job.

\ToolName{}  only mediates the \texttt{GITHUB\_TOKEN}. Other credentials available to a job, such as repository secrets, personal access tokens and OIDC \texttt{id-token}s as well as attacks on availability, are out of scope.  Defenses such as commit pinning, action immutability, and egress control are complementary. \ToolName{} also does not prevent a step from misusing the permissions it is granted, nor from tampering with state shared within the job. For instance, a malicious step can modify workspace files that a later step commits using its own \emph{contents: write} permission. \ToolName{} limits such indirect attacks to what the later step’s policy allows. This threat model is in line with GitHub's permission model and related works~\cite{argus,cosseter}, while providing finer-grained control.

\section{{\ToolName} Design}\label{sec:tool_design}

\begin{figure*}[t]
	\centering
	\scalebox{0.7}{%
		\input{./tikz/tool_design.tex}%
	}%
	\caption{{\ToolName}: Architecture and Workflow}
	\label{fig:tool_design}
\end{figure*}

{\ToolName} builds on GitHub's self-hosted runner and rests on the design decision that untrusted step code never holds the job's \texttt{GITHUB\_TOKEN}. 
Instead, each step receives a \emph{step identifier token} (SIT). 
A separate \emph{{\ToolName} controller} keeps the real token, uses the SIT to identify the step issuing each request and decides whether to forward it to GitHub. 
The runner and the controller run in separate containers. 
The steps reach the controller only through a TLS-capable proxy~\cite{mitmproxy}, and the runner's worker through a private channel. 
We present the architecture and its security goals, explain how {\ToolName}  protects the token, and finally describe how it observes and enforces step-level policies.

\subsection{Architecture and Workflow} 
Figure~\ref{fig:tool_design} shows the architecture of {\ToolName} and the workflow of a job. 
When a workflow is triggered by a repository event, GitHub assigns each job to a registered runner (Section~\ref{sec:github_actions}). 
{\ToolName}'s job listener receives the job, including its \texttt{GITHUB\_TOKEN} (\ding{182}), and starts a worker process to execute it. 
Before any step runs, the worker's \emph{Broker Manager} registers the job and its \texttt{GITHUB\_TOKEN} over the \emph{control channel} (\ding{183}) with the credential broker. 
The latter is the controller component that stores the token, and issues and revokes SITs. 
Immediately before each step, the broker manager obtains a fresh SIT bound to that step (\ding{184}) and installs it in the step's execution environment, together with the proxy configuration (\ding{185}). 
While the step runs, each request towards GitHub reaches the controller over the \emph{proxy channel}, together with the step's SIT (\ding{186}). 
The controller's \emph{Request Verifier} checks the request against the step's policy (\ding{187}) and, if accepted, replaces the SIT with the \texttt{GITHUB\_TOKEN} and forwards the request to GitHub (\ding{188}). 
When the step finishes, the broker manager has its SIT revoked by the credential broker (\ding{184}) and terminates the step's remaining processes. 
When the job terminates, the credential broker deletes the job and its token.

{\ToolName}'s architecture targets four security goals (SGs), following the threat model of Section \ref{Sub:threat_model}:
\begin{enumerate}
	\item\label{sdg:1} Preserve the confidentiality of \texttt{GITHUB\_TOKEN}.
	\item\label{sdg:2} Prevent control channel forgery.
	\item\label{sdg:3} Prevent steps from reusing another step's SIT.
	\item\label{sdg:4} Ensure that requests bypassing the {\ToolName} controller gain no access to repository resources. %
\end{enumerate}

\subsection{Protecting the Token} 
\tightpar{\texttt{GITHUB\_TOKEN} and step identifiers (SG1)} 
The unmodified runner exposes the job's \texttt{GITHUB\_TOKEN} to steps only through the \texttt{github} and \texttt{secrets} contexts.  
{\ToolName} removes the token from both as soon as the job arrives, so it reaches no step input, environment variable, or command. 
Before each step, the credential broker generates a fresh SIT and binds it to the step's identity (defined by the workflow, the job, the triggering event, the step's position in the job, and the invoked action).  
{\ToolName} runner hands the step this SIT wherever the unmodified runner would have handed it the \texttt{GITHUB\_TOKEN}, hence existing workflows and actions run unmodified. 
A SIT is valid only while its step runs. 
The steps nested in a composite action share the SIT of their enclosing top-level step, for which the step-level permissions are specified. 
The real \texttt{GITHUB\_TOKEN} exists in only two places. 
First, the controller keeps it in a separate container that steps cannot reach.
Steps of Docker actions run in containers under rootless Docker, so they cannot reach the controller either. Second, the runner's listener and worker keep it for their own requests, and both mark their memory as non-dumpable.
Hence, unlike in the attacks of Section~\ref{Sec:Motivating_example}, a step cannot read the token, even when it runs as root inside the unprivileged runner container. %
These measures together ensure SG1.

\tightpar{Control channel (SG2)}
The {\ToolName} runner and controller communicate over two channels. 
The \emph{control channel} handles the job registration and SIT management (\ding{183},\ding{184}), while the \emph{proxy channel} handles the steps' network requests. 
The worker opens the control channel before the job starts, and the credential broker binds a single session to that connection. 
It accepts SIT lifecycle requests only over this connection and rejects any other attempt to open a new session. 
Step processes neither inherit the connection nor can read the session from the worker's protected memory, which ensures SG2.

\tightpar{Step attribution (SG3)}
The worker points the step execution environment's proxy settings to the controller and sets the step's SIT as the proxy credential. 
Therefore, the controller can attribute every request to the step that issued it, and reject requests without an active SIT. 
It authorizes a request against the policy of the step that owns the proxy connection, not against any token that the request itself carries. 
Moreover, a SIT that an earlier step left behind is revoked and grants nothing, which ensures SG3.

\tightpar{Request routing (SG4)}
The worker also adds the root certificate of the controller's proxy to the trust store of the step's environment. 
Common programs such as \texttt{curl}, \texttt{gh}, and Git honor these settings, and Docker actions receive them inside their container. 
Node.js's built-in HTTP clients ignore proxy settings, hence the runner preloads a routing shim into JavaScript actions. 
The shim and the proxy settings are used only to route requests. 
They neither decide whether a request is allowed nor handle the \texttt{GITHUB\_TOKEN}, hence {\ToolName} does not rely on them for security. 
A step that bypasses them reaches GitHub with nothing but its SIT, which GitHub does not accept. 
This ensures SG4. 

\subsection{Step-level Policies}
\tightpar{Request verification}
The {\ToolName} controller inspects only connections to GitHub hosts (\ding{186}). 
The traffic to other hosts is forwarded without inspection and never receives a \texttt{GITHUB\_TOKEN}. 
For a request to GitHub, the controller first authenticates the SIT and resolves the step it was issued to, rejecting SITs that are unknown, revoked, or expired. 
It then infers the permission required by the request, \eg from its HTTP method and API path, and labels requests for which it cannot infer one as \emph{unknown}. 
Depending on the operation mode, it then records the permission or checks it against the step's policy (\ding{187}). 
A granted request is forwarded to GitHub with the \texttt{GITHUB\_TOKEN} in place of the SIT, and the response is returned to the step (\ding{188}).

\tightpar{Policies} 
A policy maps each step of a workflow, under a given triggering event, to the permissions it may use,
\ie $(W, J, C, j) \longrightarrow P$ where $W$ is the workflow, $J$ the job, $C$ the triggering event, and $j$ the step's position in the job $J$. 
Each entry also records the invoked action, so that a policy for one action is never applied to another (Listing~\ref{lst:policy_example}). 
These are the attributes to which a SIT is bound, so every request maps to exactly one entry. 
Since this includes the step's position, a structural change to a job would require observing its policy again.

\tightpar{Observation mode}
{\ToolName} is \emph{primarily} an enforcement tool, but it can also infer the permissions each step uses in its context, which addresses the context dependence of permissions (Section \ref{sec:problem_setting}). 
In observation mode, the controller enforces no policy, but it records the permission of each request for the step and forwards the request with the real token. 
Unknown requests are forwarded as well and recorded separately for review. 
Across repeated executions, {\ToolName} takes the per-scope union of the observed permissions, hence an entry only grows, \eg from \texttt{none} to \texttt{read} to \texttt{write}.
The developer must observe a workflow with trusted versions of its actions, review the observed policy, and promote it to an approved policy.

\begin{remark}
Like any dynamic analysis, {\ToolName} observes only the executed paths, so an observed policy may \textit{under-approximate} a step's permissions. 
This is deliberate, as the aim is to infer the permissions a step exercises in a \emph{specific workflow and context}, not every permission an action could ever require. Static approaches err in the opposite direction, over-approximating from code that the workflow may never reach (Section~\ref{Sub:RQ2}). 
We treat their output as an initial policy for review. 
A missing permission is not a security risk because enforcement would deny access and log it for any missed permission.
\end{remark}

\tightpar{Enforcement mode}
The request verifier looks up the approved entry for the step and checks that its recorded action matches the action the step invoked. 
It accepts a request only if the approved access level for each permission scope is at least the required one, with \texttt{none} $<$ \texttt{read} $<$ \texttt{write}. 
It rejects requests that exceed the policy, match no entry, or are unknown. 
Upon rejection, the step receives an error response that states the reason, which also appears in the workflow logs.

\begin{figure}[h]
	\centering
	\resizebox{\columnwidth}{!}{\begin{tikzpicture}[
	>={Latex[length=1.5mm]},
	exwin/.style={draw, rounded corners=1.5pt, fill=white, anchor=north west, inner sep=0pt},
	exhead/.style={fill=black!8, anchor=north west, inner sep=0pt, rounded corners=1.5pt, minimum height=0.32cm},
	extitle/.style={anchor=west, font=\bfseries\tiny, inner sep=0pt},
	excode/.style={anchor=west, inner sep=0pt, font=\ttfamily\fontsize{5.3pt}{6.3pt}\selectfont},
	exline/.style={anchor=west, inner sep=0pt, font=\fontsize{4.8pt}{5.8pt}\selectfont, align=left, text width=2.5cm},
	exflow/.style={->, thick},
	exdeny/.style={->, thick, densely dashed, color=red},
]
	\node[exwin, minimum width=4.75cm, minimum height=2.3cm] (code) at (0,0) {};
	\node[exhead, minimum width=4.75cm] at (code.north west) {};
	\node[extitle] at (0.12,-0.16) {\faNode\ step 2: \texttt{tj-actions/changed-files}};
	\node[excode] at (0.12,-0.50) {\textcolor{exComment}{// github.token and inputs hold token }\textcolor{exCap}{$\textrm{sit}_2$}};
	\node[excode] at (0.12,-0.75) {octokit = github.getOctokit(\textcolor{exCap}{$\textrm{sit}_2$})};
	\node[excode] at (0.12,-1.00) {octokit.rest.pulls.\textcolor{exKeyword}{createReview}(params)};
	\node[excode] at (0.12,-1.25) {\ \ $\hookrightarrow$ request(\{ method: \textcolor{exString}{"POST"},};
	\node[excode] at (0.12,-1.47) {\ \ \ \ \ url: \textcolor{exString}{".../pulls/\{n\}/reviews"},};
	\node[excode] at (0.12,-1.69) {\ \ \ \ \ authorization: \textcolor{exCap}{$\textrm{sit}_2$}\})};
	\node[excode] at (0.12,-1.94) {\ \ \ \ $\hookrightarrow$ \textcolor{exKeyword}{https.request}(...)};
	\node[excode] at (0.12,-2.14) {\ \ \ \ \textcolor{exComment}{// proxied using shim}};
	\node[exstepnode] at (4.4,-1.00) {1};
	\node[exstepnode] at (4.4,-1.94) {2};

	\node[exwin, fill=gray!10, minimum width=3.15cm, minimum height=2.3cm] (broker) at (5.25,0) {};
	\node[exhead, fill=black!15, minimum width=3.15cm] at (broker.north west) {};
	\node[extitle] at (5.37,-0.16) {\faNetworkWired~{\ToolName} Controller};
	\node[exline] at (5.37,-0.62) {\textcolor{exCap}{$\textrm{sit}_2$} $\rightarrow$ step~2 of \texttt{markdownlint}\\under \texttt{pull\_request}};
	\node[exline] at (5.37,-1.17) {\texttt{POST .../reviews}\\$\Rightarrow$ \texttt{pull-requests: write}};
	\node[exline] at (5.37,-1.62) {policy grants only \texttt{read}};
	\node[exline, text=red] at (5.37,-2.02) {\faTimesCircle\ 403, not allowed};
	\node[exstepnode] at (8.13,-0.6) {3};
	\node[exstepnode] at (8.13,-1.15) {4};
	\node[exstepnode] at (8.13,-2.0) {5};

	\draw[exflow] ($(code.east)+(0,-0.6)$) -- ($(code.east)+(0.15,-0.6)$) -- ($(code.east)+(0.15,0)$) -- ($(broker.west)-(0,0)$);
	\node[font=\fontsize{4.8pt}{5.8pt}\selectfont, anchor=south, inner sep=1pt]
	at ($(broker.west)-(0.2,-0.1)$) {\textcolor{exCap}{$\textrm{sit}_2$}};

	\node[draw, rounded corners=1.5pt, minimum width=3.15cm, minimum height=0.5cm, anchor=north west, font=\fontsize{4.8pt}{5.8pt}\selectfont] (github) at (5.25,-2.95) {\faGithub\ GitHub API};

	\draw[exdeny] ($(broker.west)+(0,-0.85)$) -- ($(code.east)+(0,-0.85)$);

	\draw[exdeny] (2.4,-2.3) |- (github.west);
	\node[anchor=north west, align=left, inner sep=1pt, font=\fontsize{4.8pt}{5.8pt}\selectfont, text=red] at (2.45,-2.4) {if bypass (\eg \texttt{curl} without proxy):\\only \textcolor{exCap}{$\textrm{sit}_2$} is sent not \texttt{GITHUB\_TOKEN}};
	\node[exstepnode] at (2.4,-3.2) {6};
\end{tikzpicture}}
	\caption{The attack of Listing~\ref{listing:tjchangedfiles} under {\ToolName}.}
	\label{fig:workedoutexample}\label{lst:workedoutexample}
\end{figure}

\tightpar{Example of Listing~\ref{listing:running_example}}
We assume the approved policy grants the second step of the \texttt{markdownlint} job the permission \texttt{pull-requests: read} under the \texttt{pull\_request} context (Listing~\ref{lst:policy_example}). 
The compromised \texttt{tj-actions/changed-files} of Listing~\ref{listing:tjchangedfiles} finds the SIT \textcolor{exCap}{$\textrm{sit}_2$} instead of the \texttt{GITHUB\_TOKEN} in \texttt{github.token} and its inputs (Figure~\ref{fig:workedoutexample}). 
The action calls \texttt{createReview} \exstep{1}, which delegates the call to \texttt{Octokit}'s internal request engine with $\textrm{sit}_2$ as the request credential. 
The request engine issues the request through Node.js' \texttt{https} module \exstep{2}, which the routing shim has configured to reach GitHub through the controller with $\textrm{sit}_2$ as the proxy credential. 
The controller resolves $\textrm{sit}_2$ to the second step of \texttt{markdownlint} under \texttt{pull\_request} \exstep{3} and infers that creating a review requires \texttt{pull-requests: write} \exstep{4}. 
Since the approved policy grants this step only \texttt{read} access to pull requests, the controller's request verifier rejects the request without ever providing the real \texttt{GITHUB\_TOKEN} \exstep{5}, thus preventing the attack. 
Bypassing the proxy does not help, since the request then reaches GitHub with $\textrm{sit}_2$ \exstep{6} only.

\section{Implementation}\label{sec:implementation}
{\ToolName} comprises about 1,800 lines of C\# in the runner, 3,900 lines of Python in the controller, and a 260-line JavaScript routing shim. %

\tightpar{{\ToolName} runner}%
When a job arrives, the broker manager registers the operation mode and the job with the credential broker.
The worker evaluates references to the \texttt{GITHUB\_TOKEN}, including \texttt{github.token} and \texttt{secrets.GITHUB\_TOKEN}, to a unique placeholder. 
All step types (JavaScript, composite, Docker, and run) are created through a common handler, where {\ToolName} replaces each occurrence of the placeholder in the step's evaluated inputs and environment variables with the step's SIT. 
The credential broker generates each SIT from 32 random bytes and stores only its \texttt{SHA-256} digest together with the step's identity.
The broker manager requests revocation during step finalization, hence a SIT is revoked even if its step fails or is cancelled, and then the worker terminates the step's remaining processes.
For this, the worker records the running processes before each step and, after revocation, terminates all processes started since then. 
At job completion, the credential broker deletes the \texttt{GITHUB\_TOKEN} and invalidates any remaining SITs associated with the job. 

The control channel is HTTP over a Unix-domain socket owned by the credential broker.
Since workflow steps and the worker may run under the same OS user, socket file permissions alone cannot protect this channel.
{\ToolName} therefore uses a \texttt{close-on-exec} socket descriptor to prevent step processes from inheriting the control channel, and sets a kernel flag (\texttt{PR\_SET\_DUMPABLE}) to mark the worker's memory non-dumpable.
Thus, a step starts with a SIT but without the GitHub credential.

\tightpar{Request routing}%
The worker configures the step execution environment to route requests through the {\ToolName} controller.
For proxy-aware actions, it sets the standard proxy environment variables, such as \texttt{HTTPS\_PROXY}, to the {\ToolName} controller.
It also configures the execution environment's TLS settings, including \texttt{SSL\_CERT\_FILE} and Git's \texttt{GIT\_SSL\_CAINFO}, to use a CA bundle, which is a certificate file containing both the system root certificates and the {\ToolName} controller's public CA certificate.
For actions with independent trust stores, such as Java, {\ToolName} adds the controller's CA certificate directly to their trust store.
The proxy settings direct requests to the {\ToolName} controller, while the CA bundle allows the actions to trust the TLS connections inspected by it.
Common command-line clients launched by run steps and composite actions respect these settings.
Node.js's built-in HTTP clients, however, do not honor proxy variables, so the runner preloads a small routing shim into JavaScript actions through \texttt{NODE\_OPTIONS}.
The shim installs proxy agents for the \texttt{http}, \texttt{https}, and \texttt{fetch} clients.
Docker actions and explicit job containers receive the same proxy settings and read-only access to the CA bundle.

\tightpar{{\ToolName} controller}
The {\ToolName} controller keeps the \texttt{GITHUB\_TOKEN}, CA private key, SITs, and policy files within that container.
{\ToolName} runner interacts with the controller inside this container through a \emph{control channel} and a \emph{proxy channel}.
For the control channel, the credential broker creates a random session value when the worker process initializes its Unix-socket connection.
The credential broker records both this value and the connection on which the worker initialized it, and it accepts subsequent SIT lifecycle requests only when both match, preventing unauthorized reuse of the session value.
For proxy channel, the {\ToolName} controller extends \texttt{mitmproxy} with a Python request-processing module.
Upon receiving a connection, the proxy module extracts the step SIT from the request header, verifies that it is active, and binds its digest to the proxy connection.
For every request on that connection, the module rechecks the digest against the active SIT list and thereby resolves the step that issued the request.
The {\ToolName} controller decrypts TLS traffic only for GitHub hosts and GitHub Container Registry. 
Traffic to other destinations is forwarded without inspection and never receives the \texttt{GITHUB\_TOKEN}.
For an inspected request, the request verifier validates the destination and upstream TLS connection and canonicalizes the request path. 
To infer the required permission for the request, it first matches the HTTP method and canonicalized path against $259$ GitHub REST endpoint templates, each associated with a permission. 
For example, \texttt{POST /repos/\{owner\}/\{repo\}/pulls/\{n\}/reviews} requires \texttt{pull-requests: write}.
If no template matches, the controller uses rules to infer the required permission from the resource family, which is the first path segment after the repository name, \eg \texttt{/repos/\{owner\}/\{repo\}/issues} belongs to the \texttt{issues} resource family.
We derived the initial mapping from GitHub Permissions Monitor and Advisor~\cite{GithubPer} and extended it using the GitHub REST API documentation.
GraphQL operations and requests matching neither mechanism are labeled \emph{unknown}.

We observe that different GitHub clients send the token in different ways. 
REST clients use an \texttt{Authorization} header, Git supplies the token as the password for an \texttt{x-access-token} login, Git Large File Storage (LFS) specifies its operation in the request body, and Docker exchanges its credential for a registry token.
The {\ToolName} controller therefore differentiates Git \texttt{fetch} from \texttt{push}, inspects the LFS operation, mediates the container registry's token exchange, and substitutes the real token in the format each client expects.
Some actions also keep their credential for later steps. 
For example, \texttt{actions/checkout} stores its SIT in Git configuration, so when a later command presents a SIT belonging to an earlier step, the controller verifies that it was issued within the same job but does not use it to authorize the request. 
Instead, the SIT on the connection is used to identify the current step, and the request is checked against that step's policy.
For every valid request in observation mode and every policy-approved request in enforcement mode, the broker removes all action-supplied authorization headers, swaps the SIT with the \texttt{GITHUB\_TOKEN} in the expected format, and forwards the request over a verified TLS connection to GitHub.

The {\ToolName} controller stores each workflow's observed and approved policies in separate directories on a volume that neither the runner nor any step can access.
Listing~\ref{lst:policy_example} shows the entry for the second step of the \texttt{markdownlint} job of Listing~\ref{listing:running_example}.
\definecolor{bkC1}{RGB}{214,39,40}    %
\definecolor{bkC2}{RGB}{22,22,22}     %
\definecolor{bkC3}{RGB}{44,160,44}    %
\definecolor{bkC4}{RGB}{255,127,14}   %
\definecolor{bkC5}{RGB}{148,103,189}  %
\definecolor{bkC6}{RGB}{23,190,207}   %
\definecolor{bkC7}{RGB}{31,119,180}   %
\definecolor{bkC8}{RGB}{140,86,75}    %
\definecolor{bkC9}{RGB}{227,119,194}  %
\definecolor{bkC10}{RGB}{188,189,34}  %

\newcommand{\ob}[1]{\textcolor{bkC#1}{\ttfamily\{}}
\newcommand{\cb}[1]{\textcolor{bkC#1}{\ttfamily\}}}

\lstset{
	language=json,
	breaklines=true,
	firstnumber=1,
	postbreak=\mbox{\textcolor{red}{$\hookrightarrow$}\space},
	escapeinside={(*}{*)},
}

\begin{lstlisting}[float=h, label=lst:policy_example, caption={Approved policy for the second step of \texttt{markdownlint}}, captionpos=b]
(*\ob1*) "schema_version": 1,
  "workflow_path":
    ".github/workflows/markdownlint.yml",
  "workflows": (*\ob2*)
    ".github/workflows/markdownlint.yml": (*\ob3*)
      "jobs": (*\ob4*) "markdownlint": (*\ob5*)
        "contexts": (*\ob6*) "pull_request": (*\ob7*)
          "steps": (*\ob8*) "2": (*\ob9*)
            "action": "tj-actions/changed-files@v47",
            "permissions": (*\ob{10}*) "pull-requests": "read" (*\cb{10}*)
(*\cb9*) (*\cb8*) (*\cb7*) (*\cb6*) (*\cb5*) (*\cb4*) (*\cb3*) (*\cb2*) (*\cb1*)
\end{lstlisting}

A step entry lists only the granted scopes out of the available scopes in Table~\ref{tab:github_permissions_risk}, each as \texttt{read} or \texttt{write}. 
Similar to GitHub's job-level permissions, any unlisted scope is treated as \texttt{none}.
In observation mode, the controller merges each observation per scope, records unknown requests with their method, host, and path, and saves the policy before forwarding the request.
In enforcement mode, the controller accepts a request only if a matching entry exists in the approved policy that names the invoked action and grants the required permission.
Unknown requests are rejected. 

\tightpar{Deployment}%
Our deployment runs the {\ToolName} runner, controller, and Docker-in-Docker daemon as separate Docker Compose services.
The {\ToolName} controller uses a read-only root filesystem and drops all Linux capabilities. 
Its private volume is not mounted into the runner or Docker daemon, and only the worker receives the control-socket mount.

\section{Evaluation}\label{sec:evaluations}

We evaluate  {\ToolName} by answering four research questions:
\begin{itemize}
	\item[RQ1] How prevalent are the overprivileged jobs across popular GitHub Actions workflows? 
	\item[RQ2]  Is {\ToolName} compatible with real-world workflows and how does it compare to state-of-the-art tools?
	\item[RQ3] Can {\ToolName} prevent permission misuse attacks arising from overprivileged jobs? 
	\item[RQ4] What is the performance overhead of {\ToolName}?

\end{itemize}
\tightpar{Setup} 
All our experiments  run on a machine with an AMD EPYC 7742 CPU and 32 GB of memory, running Ubuntu 22.04 LTS. 
We collected the workflows in our dataset through the GitHub REST APIs~\cite{githubRESTEndpoints} in October 2026.
\subsection{RQ1: Prevalence of Overprivileged Jobs}
\label{Sub:RQ1}
To investigate the prevalence of overprivileged jobs, we examined $500$ workflows from GitHub, selected in descending order of stars of their associated repositories, and extracted a total of $12,915$ jobs.
We used the Step Security Knowledge Base (SSKB) \cite{stepsecurityKnowledgeBase} to determine the permissions required by each GitHub Action. 
We then computed the union of permissions for all actions within each job to estimate the permissions that should be assigned to a job. We consider the job as overprivileged if the union of these permissions exceeds the permissions of at least one of its steps, i.e., if some step would have a permission that only another step needs. Actions that did not have an entry in SSKB were conservatively excluded from the analysis. 
Figure~\ref{fig:rq1_results} shows the result of this evaluation.

\begin{figure}[htbp]
\centering
\begin{tikzpicture}[
	font=\footnotesize,
	ring/.style={draw=white, line width=0.8pt},
	]
	\definecolor{RQoneMulti}{HTML}{3182BD}
	\definecolor{RQoneSingle}{HTML}{9ECAE1}
	\definecolor{RQoneIgnored}{HTML}{D9D9D9}
	\definecolor{RQoneOver}{HTML}{E6550D}
	\definecolor{RQoneNotOver}{HTML}{FDD0A2}

	\newcommand{\rqonering}[5]{%
		\fill[#1, ring] (#4:#3) arc[start angle=#4, end angle=#5, radius=#3]
			-- (#5:#2) arc[start angle=#5, end angle=#4, radius=#2] -- cycle;}

	\begin{scope}[shift={(6.0,0)}]
		\rqonering{RQoneMulti}{0.56}{1.2}{90}{-82.125}
		\rqonering{RQoneSingle}{0.56}{1.2}{-82.125}{-188.494}
		\rqonering{RQoneIgnored}{0.56}{1.2}{-188.494}{-270}
		\rqonering{RQoneOver}{1.27}{1.4}{90}{-76.857}
		\rqonering{RQoneNotOver}{1.27}{1.4}{-76.857}{-270}

		\node[align=center, font=\footnotesize\bfseries, inner sep=0pt] at (0,0) {12,915\\jobs};
		\node[font=\tiny\bfseries, text=white] at (3.94:0.88) {47.8\%};
		\node[font=\tiny\bfseries] at (-135.31:0.88) {29.5\%};
		\node[font=\tiny\bfseries] at (130.75:0.88) {22.6\%};
		\node[font=\scriptsize\bfseries, text=RQoneOver, anchor=west] at (6.57:1.44) {46.3\%};
	\end{scope}

	\newcommand{\rqoneswatch}[1]{\textcolor{#1}{\rule{7pt}{7pt}}}
	\node[anchor=west, font=\scriptsize, inner sep=0pt] at (0,0) {%
		\begin{tabular}{@{}l@{\hspace{4pt}}l@{}}
			\multicolumn{2}{@{}l}{\textbf{All jobs (inner ring)}}\\
			\rqoneswatch{RQoneMulti} & Multi-step: 6,175 (47.8\%)\\
			\rqoneswatch{RQoneSingle} & Single-step: 3,816 (29.5\%)\\
			\rqoneswatch{RQoneIgnored} & Ignored: 2,924 (22.6\%)\\[5pt]
			\multicolumn{2}{@{}l}{\textbf{Over-privilege (outer ring)}}\\
			\rqoneswatch{RQoneOver} & Over-privileged: 5,986 (46.3\%)\\
			\rqoneswatch{RQoneNotOver} & Not over-privileged: 6,929 (53.7\%)
		\end{tabular}};
\end{tikzpicture}
\caption{Prevalence of overprivileged jobs}
\label{fig:rq1_results}
\end{figure}

Our analysis divides the jobs into three categories: jobs with a single step, jobs with multiple steps, and ignored jobs. 
By definition, single-step jobs cannot be overprivileged because the job-level permission is for its single step. 
We remark that some of the $3,816$ jobs reported as single-step in Figure~\ref{fig:rq1_results} actually have multiple steps. 
However, because SSKB lacks permissions for the associated actions, they were not considered in our analysis and are reported as single-step.
Similarly, if SSKB lacks information for all the steps within a job, or the job does not use any marketplace actions (\eg it uses the \texttt{run} command or local actions), we report it as ignored.

For jobs with multiple steps, we extracted the permissions for each step from SSKB. 
Our analysis revealed that out of $6,175$ jobs with multiple steps, $5,986$ were identified as overprivileged, representing approximately $46.3\%$ of the total jobs. 
These results motivate the need for enforcing fine-grained permission, as provided by {\ToolName}.

\tightpar{Choice of dataset}
We also considered the permission dataset released by \textsc{Cosseter}~\cite{cosseter_dataset} as the basis of this evaluation.
However, it provides complete permission data for only a small subset of the actions used in our workflows, substantially reducing the coverage: $56.1\%$ of the jobs were ignored and only $1,867$ jobs ($14.5\%$) had multiple steps with known permissions, compared to $22.3\%$ and $6,212$ jobs ($48.1\%$) for SSKB. 
Thus, we chose SSKB for this analysis.

\subsection{RQ2: Compatibility and SotA Comparison}\label{Sub:RQ2}

We first check that {\ToolName} does not change the behavior of real-world workflows due to its instrumentation.
We then validate  \ToolName's  observed permissions against GitHub Advisor~\cite{GithubPer}, a dynamic monitor that reports job-level permissions. 
Finally, we compare its context-dependent, step-level permission observations with action-level permissions of SSKB~\cite{stepsecurityKnowledgeBase} and \textsc{COSSETER}~\cite{cosseter}.

\tightpar{Job selection} Dynamic tools like {\ToolName} and GitHub Advisor must run workflows end to end, which requires configuring their triggers, inputs, and dependencies, some relying on third-party services. Tools that mock these missing parameters and dependencies, such as ACT~\cite{act}, failed on many of our target workflows and repositories. Moreover, selecting jobs solely by repository stars (as we did in RQ1) would mostly repeat the same few actions, limiting the coverage and diversity of the evaluation. We therefore select jobs from the overprivileged jobs of RQ1 with the goal of maximizing the diversity of actions, after excluding jobs that need third-party secrets or non-Linux runners, as detailed in Appendix~\ref{appx:RQ2selection}. These selection criteria yielded $48$ jobs across $46$ repositories, which use $127$ distinct third-party actions of which $92$ were JavaScript, $26$ composite, and $9$ Docker actions.
This experiment uses the observation mode of {\ToolName}, which is context-dependent. 

\subsubsection{Compatibility}%
To evaluate the compatibility of {\ToolName}, we execute the workflows of target jobs once with the unmodified GitHub runner and again with {\ToolName} in observation mode, and compare the results of each job and of each of its steps.

To run the workflows, we import the latest commit of the default branch of each repository associated with each workflow into our GitHub organization and manually analyze the workflows to identify all specified events and execution conditions. 
Then, we wrote a set of execution scripts for each mode using GitHub CLI tool~\cite{githubcli} to trigger the events and satisfy the required conditions for all target jobs.
All inputs to the workflows (\eg issue body for issue event), are kept identical for all workflow runs to ensure that the execution context is consistent across the runs.
If a job is triggered by multiple events, we execute the workflow for each event separately to ensure all contexts are covered.

\begin{figure}[ht]
	\centering
	\definecolor{RQtwoSucceeded}{HTML}{3182BD}
	\definecolor{RQtwoToolFailure}{HTML}{E6550D}
	\definecolor{RQtwoUpstream}{HTML}{D9D9D9}
	\scalebox{0.9}{%
	\begin{tikzpicture}
		\begin{axis}[
			xbar stacked, scale only axis,
			width=0.72\linewidth, height=1.5cm,
			bar width=11pt,
			xmin=0, xmax=48, ymin=-0.6, ymax=1.6,
			ytick={0,1}, yticklabels={\textbf{{\ToolName}}, Unmodified Runner},
			xtick={0,8,16,24,32,40,48},
			tick label style={font=\footnotesize},
			xlabel={Jobs}, xlabel style={font=\footnotesize},
			legend style={font=\scriptsize, draw=none, at={(0.5,-0.78)}, anchor=north, legend columns=3,
				/tikz/every even column/.append style={column sep=6pt}},
			legend cell align=left,
			legend image code/.code={\draw[#1] (0cm,-0.08cm) rectangle (0.32cm,0.08cm);},
		]
			\addplot[draw=black, line width=0.3pt, fill=RQtwoSucceeded] coordinates {(40,1) (39,0)};
			\addlegendentry{Succeeded}
			\addplot[draw=black, line width=0.3pt, fill=RQtwoToolFailure] coordinates {(0,1) (1,0)};
			\addlegendentry{Tool failure}
			\addplot[draw=black, line width=0.3pt, fill=RQtwoUpstream] coordinates {(8,1) (8,0)};
			\addlegendentry{Failed upstream or drifted}
			\coordinate (rqo0) at (axis cs:20,1);
			\coordinate (rqo1) at (axis cs:44,1);
			\coordinate (rqo5) at (axis cs:19.5,0);
			\coordinate (rqo6) at (axis cs:44,0);
		\end{axis}
		\node[font=\scriptsize\bfseries, text=white, inner sep=1pt] at (rqo0) {40};
		\node[font=\scriptsize\bfseries, text=black, inner sep=1pt] at (rqo1) {8};
		\node[font=\scriptsize\bfseries, text=white, inner sep=1pt] at (rqo5) {39};
		\node[font=\scriptsize\bfseries, text=black, inner sep=1pt] at (rqo6) {8};
	\end{tikzpicture}}
	\caption{{\ToolName} Compatibility comparison results}
	\label{fig:rq2_compatibility}
\end{figure}
As reported in Figure~\ref{fig:rq2_compatibility}, $39$ of the $48$ jobs executed successfully on both GitHub runner and {\ToolName}.
After manually inspecting the execution logs, we found that $8$ jobs failed to execute with both tools due to upstream failures or drifted dependencies. 
For example, the ApplyFlow repository was using a removed Docker image (minio/minio) as part of its workflow.
The remaining job, which executed successfully on GitHub runner but failed on {\ToolName},
uses a Docker action (\emph{Google's OSS-Fuzz}) that builds a Docker image through the host's Docker socket. 
To avoid privilege escalation attacks, {\ToolName}  does not expose this socket by design because a step could use it to escape its container and reach the controller,  thus violating SG1. 

\subsubsection{{\ToolName} vs SotA tools}%
In this experiment, we compare the permissions observed by {\ToolName} with those reported by the state-of-the-art tools.

\tightpar{Granite vs GitHub Advisor}
We first compare {\ToolName} with GitHub Advisor~\cite{GithubPer}, an independent dynamic monitor that reports permissions at job level.
We execute each workflow with the same context and inputs for both GitHub Advisor and {\ToolName} in observation mode. 
Because GitHub Advisor reports permissions at job level and {\ToolName} operates at the step-level, we aggregate {\ToolName}'s observed step-level permissions for each job before comparing the results.
After analyzing the execution logs, GitHub Advisor failed to execute $11$ of the jobs that  {\ToolName} succeeds,  because of three main reasons: 
(1) $7$ jobs lose their connection to GitHub Actions Service after running for one hour, causing the runners to stop reporting their status to the service. 
Since GitHub Advisor intercepts all network traffic, including the communication between the runner and the GitHub service, this may have contributed to the connection loss.
(2) $3$ jobs exceed the timeout limit set in their workflow file, and
(3) $1$ job fails due to a malformed header response from a non-GitHub website.
Intercepting all network communication increases the execution time substantially and may result in unsupported response headers (as observed with HTTP/2 in the last case), likely causing the job failures.
Although {\ToolName} successfully executed all $11$ jobs, we exclude these jobs from the comparison because GitHub Advisor failed to execute them and to report their required permissions.
The results of the job-level permission comparison are summarized in Figure~\ref{fig:rq2_advisor}.
\begin{figure}[ht]
	\centering
	\definecolor{RQtwoAdvExact}{HTML}{3182BD}
	\definecolor{RQtwoAdvPublic}{HTML}{9ECAE1}
	\definecolor{RQtwoAdvMismatch}{HTML}{E6550D}
	\scalebox{0.9}{%
	\begin{tikzpicture}
		\begin{axis}[
			xbar stacked, scale only axis,
			width=0.9\linewidth, height=0.7cm,
			bar width=12pt,
			xmin=0, xmax=30, ymin=-0.5, ymax=0.5,
			ytick=\empty,
			xtick={0,5,10,15,20,25,30},
			tick label style={font=\footnotesize},
			xlabel={Executed jobs}, xlabel style={font=\footnotesize},
			legend style={font=\scriptsize, draw=none, at={(0.5,-1.55)}, anchor=north, legend columns=3,
				/tikz/every even column/.append style={column sep=8pt}},
			legend cell align=left,
			legend image code/.code={\draw[#1] (0cm,-0.08cm) rectangle (0.32cm,0.08cm);},
		]
			\addplot[draw=black, line width=0.3pt, fill=RQtwoAdvExact] coordinates {(4,0)};
			\addlegendentry{Exact match}
			\addplot[draw=black, line width=0.3pt, fill=RQtwoAdvPublic] coordinates {(23,0)};
			\addlegendentry{excluded public-data match}
			\addplot[draw=black, line width=0.3pt, fill=RQtwoAdvMismatch] coordinates {(3,0)};
			\addlegendentry{Mismatch}
			\coordinate (rqa0) at (axis cs:2.0,0);
			\coordinate (rqa1) at (axis cs:15.5,0);
			\coordinate (rqa2) at (axis cs:28.5,0);
		\end{axis}
		\node[font=\scriptsize\bfseries, text=white, inner sep=1pt] at (rqa0) {4};
		\node[font=\scriptsize\bfseries, text=black, inner sep=1pt] at (rqa1) {23};
		\node[font=\scriptsize\bfseries, text=white, inner sep=1pt] at (rqa2) {3};
	\end{tikzpicture}}
	\caption{Job-level permissions observed by {\ToolName} and GitHub Advisor}
	\label{fig:rq2_advisor}
\end{figure}

The main reason for the permission discrepancies in $23$ jobs is that GitHub Advisor ignores permissions for API calls that access public data (\eg \texttt{action:read} on a public repository). 
In contrast, {\ToolName} observes \emph{all} permissions given that permission enforcement is its primary objective.
After excluding public data reads from the permissions reported by {\ToolName}, job level permissions agree on $27$ jobs and differ in $3$ cases.
The mismatch cases are detailed in Appendix~\ref{appx:rq2_comparison}. 
Briefly, the main reasons for the mismatch cases are:
(1) Two jobs pushed images to the container registry (\texttt{ghcr.io}), requiring the \texttt{packages:write} permission that GitHub Advisor failed to report because it does not support requests to this domain.
(2) GitHub Advisor does not support Docker actions, so in the case of \texttt{EnricoMi/publish-unit-test-result-action}, the \texttt{checks: write} permission was missed.
We believe these two cases are a limitation of GitHub Advisor and demonstrate the advantages of {\ToolName} over GitHub Advisor.

\tightpar{Overprivileged jobs}%
We also analyze the step-level permissions observed by {\ToolName} to identify overprivileged jobs.  
Recall from Section~\ref{sec:problem} that under GitHub's current permission model, every step receives all permissions assigned to its job.
We therefore classify a job as overprivileged if at least one of its steps does not use all permissions assigned to that job.
Of the $39$ jobs executed with {\ToolName} in observation mode, $38$ (approximately $97\%$) are overprivileged. This confirms at runtime the estimate of RQ1, from which these jobs were drawn.
Moreover, excluding the three read permissions (\ie \texttt{contents:read}, \texttt{actions:read}, and \texttt{attestations:read}) that are public on a public repository reduces this number to $14$ overprivileged jobs (approximately $35\%$).

\tightpar{{\ToolName} vs \textsc{Cosseter}}
We compare the results of {\ToolName} with \textsc{Cosseter}, which is a static analysis tool to extract action permissions.
The $39$ jobs executed with {\ToolName} contain $178$ (job, action) pairs. 
For each pair, we compare the permissions observed by {\ToolName} for the action in that job with the permissions reported by \textsc{Cosseter} for that action. 
Figure~\ref{fig:rq2_static_comparison} summarizes the results across all $178$ pairs.
\textsc{Cosseter} dataset does not report any permissions for 132 pairs of (job, action) since it does not cover the corresponding 83 actions, hence we exclude these pairs from the comparison.
We say that a \emph{permission addition} occurs when \textsc{Cosseter} lists a permission that was either not observed by {\ToolName} or has a higher access level than the one observed.
Conversely, a \emph{permission omission} occurs when {\ToolName} observes a permission that \textsc{Cosseter} did not list, or it has a higher access level than listed.
Our analysis identifies $10$ pairs with permission additions, caused by $5$ distinct actions, and $3$ pairs with permission omissions, involving $3$ distinct actions. 
The remaining $33$ pairs report exactly matching permissions.
Permission additions may correspond to possible action paths that were not exercised under our evaluated contexts and therefore do not imply that those permissions are unnecessary in every execution. 
In contrast, permission omissions identify permissions that were exercised during evaluation but absent from \textsc{Cosseter}'s output. 
We believe this is due to:
(1) Static analysis of the action's source code misses dynamic imports and child processes for API calls,
(2) Indirect API calls, where the API method is passed to another function, \eg Octokit's helper, instead of being called directly, so the analysis does not identify the request.
For example, two actions \eg \texttt{ad-m/github-push-action} push a commit by running \textsc{git} CLI as a child process. 
\textsc{Cosseter} sees a command execution rather than a GitHub REST API call, therefore misses the required \texttt{contents:write} permission.
The other action, \texttt{dorny/paths-filter}, uses an indirect API call only for pull-request events, so \textsc{Cosseter} misses the required \texttt{pull-requests:read} permission.
We manually inspected the source code of the relevant actions in the $10$ job-level permission addition cases to determine whether the additional permissions could be exercised in the same execution context by varying runtime input values. 
In six pairs, \texttt{actions/upload-artifact} uploads artifacts with a separate token instead of \texttt{GITHUB\_TOKEN}, so the \texttt{actions:write} permission listed by \textsc{Cosseter} is never required.
In the remaining four pairs, the relevant paths depend on the execution context, \eg \texttt{peter-evans/create-pull-request} opens a pull request only if the job modified files. In 8 cases, \textsc{Cosseter} reports write permissions that are either not needed by action or never exercised during the corresponding executions.
The details of the permission omissions/addition are reported in Appendix~\ref{appx:rq2_cosseter_omissions}.

\begin{figure}[t]
	\centering
	\definecolor{RQtwoExact}{HTML}{3182BD}
	\definecolor{RQtwoAdd}{HTML}{FDAE6B}
	\definecolor{RQtwoOmit}{HTML}{E6550D}
	\definecolor{RQtwoBoth}{HTML}{8C2D04}
	\definecolor{RQtwoNone}{HTML}{D9D9D9}
	\begin{tikzpicture}[baseline={(0,0)}]
		\fill[RQtwoExact, draw=white, line width=0.6pt] (0,0) -- (90.000:1.25) arc[start angle=90.000, end angle=-122.360, radius=1.25] -- cycle;
		\node[font=\scriptsize\bfseries] at (-16.180:0.8) {105};
		\fill[RQtwoAdd, draw=white, line width=0.6pt] (0,0) -- (-122.360:1.25) arc[start angle=-122.360, end angle=-138.539, radius=1.25] -- cycle;
		\node[font=\scriptsize\bfseries, inner sep=0pt] at (-130.449:0.97) {8};
		\fill[RQtwoOmit, draw=white, line width=0.6pt] (0,0) -- (-138.539:1.25) arc[start angle=-138.539, end angle=-152.697, radius=1.25] -- cycle;
		\node[font=\scriptsize\bfseries, anchor=north east, inner sep=1pt] at (-145.618:1.3) {7};
		\fill[RQtwoBoth, draw=white, line width=0.6pt] (0,0) -- (-152.697:1.25) arc[start angle=-152.697, end angle=-158.764, radius=1.25] -- cycle;
		\node[font=\scriptsize\bfseries, anchor=south east, inner sep=1pt] at (-155.730:1.3) {3};
		\fill[RQtwoNone, draw=white, line width=0.6pt] (0,0) -- (-158.764:1.25) arc[start angle=-158.764, end angle=-270.000, radius=1.25] -- cycle;
		\node[font=\scriptsize\bfseries] at (-214.382:0.8) {55};
		\node[font=\small\bfseries] at (0,1.65) {SSKB};
	\end{tikzpicture}
	\hspace{0.4cm}
	\begin{tikzpicture}[baseline={(0,0)}]
		\fill[RQtwoExact, draw=white, line width=0.6pt] (0,0) -- (90.000:1.25) arc[start angle=90.000, end angle=23.258, radius=1.25] -- cycle;
		\node[font=\scriptsize\bfseries] at (56.629:0.8) {33};
		\fill[RQtwoAdd, draw=white, line width=0.6pt] (0,0) -- (23.258:1.25) arc[start angle=23.258, end angle=3.034, radius=1.25] -- cycle;
		\node[font=\scriptsize\bfseries, inner sep=0pt] at (13.146:0.97) {10};
		\fill[RQtwoOmit, draw=white, line width=0.6pt] (0,0) -- (3.034:1.25) arc[start angle=3.034, end angle=-3.034, radius=1.25] -- cycle;
		\node[font=\scriptsize\bfseries, anchor=west] at (0.000:1.32) {3};
		\fill[RQtwoNone, draw=white, line width=0.6pt] (0,0) -- (-3.034:1.25) arc[start angle=-3.034, end angle=-270.000, radius=1.25] -- cycle;
		\node[font=\scriptsize\bfseries] at (-136.517:0.8) {132};
		\node[font=\small\bfseries] at (0,1.65) {\textsc{Cosseter}};
	\end{tikzpicture}

	\medskip
	{\scriptsize \textcolor{RQtwoExact}{\rule{7pt}{7pt}}~Exact match \quad \textcolor{RQtwoAdd}{\rule{7pt}{7pt}}~Addition only \quad \textcolor{RQtwoOmit}{\rule{7pt}{7pt}}~Omission only \\[2pt]
	\textcolor{RQtwoBoth}{\rule{7pt}{7pt}}~Addition and omission \quad \textcolor{RQtwoNone}{\rule{7pt}{7pt}}~Not covered by the tool}
	\caption{{\ToolName} comparison with Cosseter and SSKB}
	\label{fig:rq2_static_comparison}
\end{figure}

\tightpar{Granite vs SSKB}
We compare {\ToolName} with SSKB using the same definitions of permission additions and omissions and the same process as \textsc{Cosseter}.
The results are reported in Figure~\ref{fig:rq2_static_comparison}.
We identify $8$ permission additions, caused by $8$ distinct actions, and $7$ omissions, involving $7$ distinct actions, while three pairs show both.
In $105$ pairs the permissions match, and SSKB does not report any permissions for the remaining $55$ pairs, which we exclude from the comparison, as for \textsc{Cosseter}.
Hence, action-level permissions are both incomplete, covering only $123$ and $46$ of the $178$ pairs, and imprecise for a given workflow. Additions would leave steps overprivileged, with unneeded scopes, while omissions would break their jobs if enforced.
These results show that workflows frequently exercise only a subset of the permissions associated with an action. Actions can support multiple behaviors, whereas a particular workflow invokes them under specific triggering events, within their configured workflow.
{\ToolName} uses this context to produce context-dependent step-level policies.

\subsection{RQ3: Security Evaluation} \label{Sub:RQ3}
We evaluate the security benefits of {\ToolName} by conducting three experiments: 
(1) assessment of the severity of overprivileged jobs;
(2) implementation and evaluation of attacks that exploit the permissions of overprivileged jobs; and 
(3) measurement of attack surface reduction with respect to GitHub's current permission model.
For these experiments, we select the $20$ most popular repositories  with jobs that RQ1 identified as overprivileged  and execute them with {\ToolName} in observation mode. %
The analysis confirms $42$ overprivileged jobs across these $20$ repositories. 
We use {\ToolName}'s observation mode for severity assessment, and enforcement mode for attack prevention and attack surface reduction. %

\tightpar{Severity assessment}%
We assess the severity of the $42$ overprivileged jobs according to the risk assessment criteria of Table~\ref{tab:github_permissions_risk}, with the results summarized in Figure~\ref{fig:rq2_severity}.
\begin{figure}[ht]
	\centering
	\begin{tikzpicture}[scale=0.64]
		\pie[
		radius=1.5,
		text=pin,
		pos={0,0},
		hide number,
		font=\scriptsize,
		color={contentsColor!70, pullRequestsColor!70, securityEventsColor!70, checksColor!70, pullRequestsColor!70, contentsColor!70, issuesColor!70, actionsColor!70},
		style={piechart},
		rotate=90,
		sum=83
		]{
			20/contents (20),
			19/pull-requests (19),
			10/security-events (10),
			1/checks (1),
			6/pull-requests (6),
			17/contents (17),
			2/issues (2),
			8/actions (8)
		}
		
		\pie[
		radius=1.05,
		hide number,
		style={piechart},
		color={white},
		sum=83
		]{
			83/
		}
		
		\pie[
		radius=1,
		text=legend,
		color={AwesomeRed, yellow, cyan, MediumSeaGreen},
		style={piechart},
		font=\scriptsize,
		rotate=90,
		sum=83
		]{
			20/Critical,
			29/High,
			1/Medium,
			33/Low
		}
	\end{tikzpicture}
	\caption{Severity of overprivileged scopes}
	\label{fig:rq2_severity}
\end{figure}

We find $83$ different types of overprivileged permission scopes across the $42$ jobs. These are the permissions of a job that at least one of its steps does not use. Some jobs contain multiple overprivileged scopes due to different steps within the same job, with an average of $1.9$ overprivileged scopes per job. 
The results show  that $20$ ($24.1\%$) instances are classified as Critical in Table~\ref{tab:github_permissions_risk}, $29$ ($34.9\%$) as High, $1$ ($1.2\%$) as Medium, and $33$ ($39.8\%$) as Low severity. 
This indicates that a significant portion of the overprivileged scopes ($59.0\%$) have High or Critical severity levels, with the most frequent overprivileged scopes being \texttt{contents} ($37$ cases), \texttt{pull-requests} ($25$ cases), \texttt{security-events} ($10$ cases), \texttt{actions} ($8$ cases), \texttt{issues} ($2$ cases), and \texttt{checks} ($1$ case). 
Since Table~\ref{tab:github_permissions_risk} rates risks for a public repository, this is a lower bound. %

\begin{table*}[ht]
	\centering
	\caption{Security Evaluations of Compromised GitHub Actions}
	\label{tab:security_analysis}
	\scalefont{0.75}{
		\begin{tabular}{l l l c l c}
			\textbf{Repository Name} & \textbf{Workflow Name} & \textbf{Threat Action} & \textbf{Overprivileged Scope} & \textbf{Attack Type} & \textbf{Technique}\\
			\midrule[1px]
			Significant-Gravitas/AutoGPT & classic-benchmark-publish-package.yml & actions/checkout@v4 & \cellcolor{red!10}contents:write & dependency substitution & T1 \\
			\rowcolor{gray!10} Significant-Gravitas/AutoGPT & classic-frontend-ci.yml & subosito/flutter-action@v2 & \cellcolor{red!10}contents:write & environment variable exfiltration & T1 \\
			axios/axios & pr.yml & actions/checkout@v4 & \cellcolor{red!10}contents:write & backdoor injection & T3 \\
			\rowcolor{gray!10} axios/axios & Publish.yml & ffurrer2/extract-release-notes@v2 & \cellcolor{red!10}contents:write & unit test bypass & T4 \\
			f/awesome-chatgpt-prompts & main.yml & ruby/setup-ruby@v1 & \cellcolor{red!10}contents:write & prompts modification & T0 \\
			\rowcolor{gray!10} langgenius/dify & translate-i18n-base-on-english.yml & pnpm/action-setup@v4 & \cellcolor{red!10}contents:write & LLM api key exfiltration & T2 \\
			mrdoob/three.js & report-size.yml & peter-evans/find-comment@v3 & \cellcolor{orange!10}pull-requests:write & approve and merge pull-request & T0 \\
			\rowcolor{gray!10} nodejs/node & scorecard.yml & actions/upload-artifact@v1 & \cellcolor{orange!10}security-events:write & dismiss security alerts & T4 \\
			twbs/bootstrap & js.yml & actions/setup-node@v4 & \cellcolor{yellow!10}checks:write & attach fake security checks & T3 \\
			\rowcolor{gray!10} jwasham/coding-interview-university & links-checker.yml & micalevisk/last-issue-action & \cellcolor{green!10}issues:write & phishing via issue comments & T0 \\
	\end{tabular}}
\end{table*}

\tightpar{Attack implementation and evaluation}
To assess the effectiveness of {\ToolName} in preventing permission misuse attacks, we select $10$ overprivileged jobs from $8$ different repositories, and implement a series of attacks that exploit the permissions granted to overprivileged jobs.
Specifically, we choose $6$ workflows with Critical severity, $2$ with High severity, $1$ with Medium severity, and $1$ with Low severity. 
Table~\ref{tab:security_analysis} details the selected workflows, including the repository name, the workflow name, the targeted action within the overprivileged job (threat action), the overprivileged scope, the type of attack, and the technique that could have been used for an attempted  bypass of {\ToolName}. 
The attack types are selected based on the repository context and the overprivileged scope. 
For example, we use \texttt{Significant-Gravitas{\slash}AutoGPT}, an open-source AI agent framework, to exploit the overprivileged \texttt{contents: write} scope to perform dependency substitution by modifying the contents of a package file in the repository.

To implement these attacks, we \emph{import} the corresponding threat action repository, modify the action's source code to include the malicious payload, and push the modified source code to the \emph{imported} repository and point the original workflow to the same repository.
The modified action thus executes the malicious payload within the original workflow's job context.
Each malicious payload exploits the overprivileged scope, either directly through the job token it receives (T0), or after trying to bypass the least-privilege enforcement of {\ToolName} using one of the following techniques, each targeting one of the security design goals in Section~\ref{sec:tool_design}:
(T1) reading the job's \texttt{GITHUB\_TOKEN} from the workflow contexts, environment variables and secrets~(security goal~(\ref{sdg:1}));
(T2) connecting to the credential broker to open a control channel and request a new SIT~(security goal~(\ref{sdg:2}));
(T3) reusing a SIT from a process that writes its SIT on disk~(security goal~(\ref{sdg:3})), \eg \texttt{actions/checkout} written credential to git config; or
(T4) bypassing the proxy and sending requests directly to the GitHub API~(security goal~(\ref{sdg:4})).
Appendix~\ref{appx:attack_implementation} details the techniques used to bypass {\ToolName}'s  least-privilege enforcement.
A modified action executes the malicious payload within the original workflow's job context and misuses the job-level permissions given by GitHub's current permission model. 

We first execute each attack with the unmodified runner and verify its success by checking its effect on the target repository.
After confirming the attack, we then re-execute the workflow with {\ToolName} in enforcement mode, using the policy promoted from the permissions observed by {\ToolName} in severity assessment experiment. As expected, {\ToolName} successfully prevents all implemented attacks. 
T0 requests are denied due to least-privilege enforcement, while requests combined with T1-T4 are denied due to both the security design goals of {\ToolName} and its least-privilege enforcement. T1 and T3 obtain only a SIT, whose requests are checked against the permissions of the step; T2 requests are denied because the credential broker rejects the request for a new SIT, and T4 requests are denied because the SIT is not valid for GitHub.
The variety of attacks implemented in this section shows the risk of overprivileged jobs to the security of CI/CD pipelines. 
{\ToolName}  effectively mitigates these threats, reducing the attack surface and enhancing the security of GitHub workflows. %

\tightpar{Attack surface reduction}
Finally, we evaluate {\ToolName}'s effectiveness in reducing the attack surface relative to the current job-level permissions model. To this end, we analyze overprivileged jobs before and after applying {\ToolName}'s enforcement, measuring  how many steps lose \texttt{write} access under Granite.
The results of this analysis are reported in Table~\ref{tab:attack_surface_reduction} in Appendix~\ref{appx:attack_reduction}.

For each overprivileged job, we analyzed the number of steps requiring \texttt{write} permissions on any scope relative to the total number of steps that have access to that scope, under the current job-level permission model. 
In the current model, each overprivileged job, with an average of $6$ steps per job, will be given \texttt{write} access, while on average only $1$ step needs that permission. 
This shows that, on average, {\ToolName} can reduce the number of steps with \texttt{write} access by approximately $83.3\%$ per overprivileged job, illustrating the effectiveness of {\ToolName} in minimizing the attack surface of GitHub Actions workflows.

\subsection{RQ4: Performance Overhead}
\label{sec:performance_overhead}
This section reports the performance overhead of {\ToolName}. %
For this evaluation, we selected $10$ most frequently used JavaScript actions, $5$ composite actions, and $1$ Docker actions from the analysis in RQ2 (Section~\ref{Sub:RQ2}). 
As workflow execution heavily depends on the repository context (\eg the size of the repository), we created a controlled workflow within an open-source project~\cite{pythontelegrambot} with $12$ distinct jobs to ensure consistent and comparable measurements across all selected actions. 
Certain actions require specific preparation, such as installing dependencies or configuring the environment. 
Therefore, additional steps were included within the jobs to accommodate these requirements.
The complete structure of this workflow, including the actions within each step, is provided in Figure~\ref{fig:performance_workflow} in Appendix~\ref{appx:performance}.
The illustrated workflow is triggered by four different events: \texttt{push}, \texttt{pull-request}, \texttt{release}, and \texttt{workflow-dispatch}. 
We trigger each event five times and measure the execution time of each step within the jobs with the unmodified runner, {\ToolName} in enforcement mode, and {\ToolName} in observation mode.

For every step, we record the time taken to execute the step from start to finish, including SIT activation/deactivation and step execution itself.
For composite actions, execution time also includes the execution of their embedded steps.
Some steps are executed under multiple triggers and may comprise \texttt{pre}, \texttt{main}, and \texttt{post} stages. 
Accordingly, we consider each unique combination of step, stage, and trigger as a separate observation, resulting in $154$ matched observations.
For each observation, we take the average of the five recorded executions and for each step calculate the overhead introduced by {\ToolName} relative to the unmodified runner baseline.
To ensure that the measurements are not influenced by external factors such as system caching, we reset the cache directory before each workflow execution. 
The summary results of this analysis are illustrated in Figure~\ref{fig:rq4_overhead_steptype} in Appendix~\ref{appx:performance}.

\begin{figure}[t]
	\centering
	\pgfplotsset{
		compat=1.18,
		rq4bar/.style={draw=black, line width=0.3pt},
		rq4legend/.style={legend image code/.code={\draw[##1] (0cm,-0.08cm) rectangle (0.32cm,0.08cm);}},
	}
	{\centering\scriptsize\ref*{rq4legendA}\par}\smallskip
	\begin{tikzpicture}[trim axis left, trim axis right]
		\begin{axis}[
			xbar stacked, scale only axis,
			width=4.9cm, height=4.6cm,
			bar width=6.5pt,
			xmin=0, xmax=100.5, ymin=-0.6, ymax=4.6,
			clip=false,
			nodes near coords, point meta=explicit symbolic, nodes near coords style={font=\fontsize{6}{7}\selectfont, inner sep=0.5pt},
			ytick={0,1,2,3,4},
			yticklabels={Docker action, Composite, JavaScript, Run, \textbf{All steps}},
			yticklabel style={xshift=-4pt, text width=1.4cm, align=right, font=\scriptsize},
			ytick style={draw=none},
			tick label style={font=\footnotesize},
			xlabel={Share of the overhead (\%)},
			xlabel style={font=\footnotesize},
			xtick={0,25,50,75,100},
			title={\textbf{Where the overhead comes from}},
			title style={font=\small},
			rq4legend,
			legend to name=rq4legendA,
			legend style={font=\footnotesize, draw=none, legend columns=2, /tikz/every even column/.append style={column sep=8pt}},
			legend cell align=left,
		]
			\addplot[rq4bar, fill=orange!80!black, every node near coord/.append style={text=white}] coordinates {
				(64.3,4.2) [64.3] (11.3,3.2) [11.3] (25.0,2.2) [25.0] (93.0,1.2) [93.0] (77.9,0.2) [77.9]};
			\addlegendentry{Policy write}
			\addplot[rq4bar, fill=blue!60!black, every node near coord/.append style={text=white}] coordinates {
				(0.6,4.2) [{}] (0.3,3.2) [{}] (0.5,2.2) [{}] (0.7,1.2) [{}] (1.9,0.2) [{}]};
			\addlegendentry{Request verification}
			\addplot[rq4bar, fill=cyan!45, every node near coord/.append style={text=black}] coordinates {
				(1.9,4.2) [{}] (5.3,3.2) [{}] (2.3,2.2) [{}] (1.3,1.2) [{}] (14.5,0.2) [14.5]};
			\addlegendentry{TLS interception}
			\addplot[rq4bar, fill=green!55!black, every node near coord/.append style={text=white}] coordinates {
				(32.2,4.2) [32.2] (80.8,3.2) [80.8] (69.7,2.2) [69.7] (5.0,1.2) [{}] (4.6,0.2) [{}]};
			\addlegendentry{Proxy relay and other}
			\addplot[rq4bar, fill=gray!45, every node near coord/.append style={text=black}] coordinates {
				(1.0,4.2) [{}] (2.2,3.2) [{}] (2.4,2.2) [{}] (0.1,1.2) [{}] (1.1,0.2) [{}]};
			\addlegendentry{Control channel communication}
			\node[anchor=east, font=\tiny, text=black!65, inner sep=1pt] at (axis cs:0,4.2) {O};
			\node[anchor=east, font=\tiny, text=black!65, inner sep=1pt] at (axis cs:0,3.8) {E};
			\node[anchor=east, font=\tiny, text=black!65, inner sep=1pt] at (axis cs:0,3.2) {O};
			\node[anchor=east, font=\tiny, text=black!65, inner sep=1pt] at (axis cs:0,2.8) {E};
			\node[anchor=east, font=\tiny, text=black!65, inner sep=1pt] at (axis cs:0,2.2) {O};
			\node[anchor=east, font=\tiny, text=black!65, inner sep=1pt] at (axis cs:0,1.8) {E};
			\node[anchor=east, font=\tiny, text=black!65, inner sep=1pt] at (axis cs:0,1.2) {O};
			\node[anchor=east, font=\tiny, text=black!65, inner sep=1pt] at (axis cs:0,0.8) {E};
			\node[anchor=east, font=\tiny, text=black!65, inner sep=1pt] at (axis cs:0,0.2) {O};
			\node[anchor=east, font=\tiny, text=black!65, inner sep=1pt] at (axis cs:0,-0.2) {E};
		\end{axis}
		\begin{axis}[
			xbar stacked, scale only axis,
			width=4.9cm, height=4.6cm,
			bar width=6.5pt,
			xmin=0, xmax=100.5, ymin=-0.6, ymax=4.6,
			clip=false,
			nodes near coords, point meta=explicit symbolic, nodes near coords style={font=\fontsize{6}{7}\selectfont, inner sep=0.5pt},
			axis lines=none, ticks=none,
		]
			\addplot[rq4bar, fill=blue!25, every node near coord/.append style={text=black}] coordinates {
				(1.6,3.8) [{}] (0.3,2.8) [{}] (0.7,1.8) [{}] (9.4,0.8) [9.4] (8.4,-0.2) [8.4]};
			\addplot[rq4bar, fill=cyan!15, every node near coord/.append style={text=black}] coordinates {
				(5.4,3.8) [{}] (6.0,2.8) [6.0] (3.1,1.8) [{}] (18.2,0.8) [18.2] (65.6,-0.2) [65.6]};
			\addplot[rq4bar, fill=green!30, every node near coord/.append style={text=black}] coordinates {
				(90.1,3.8) [90.1] (91.2,2.8) [91.2] (93.0,1.8) [93.0] (71.6,0.8) [71.6] (20.9,-0.2) [20.9]};
			\addplot[rq4bar, fill=gray!15, every node near coord/.append style={text=black}] coordinates {
				(2.8,3.8) [{}] (2.5,2.8) [{}] (3.2,1.8) [{}] (0.8,0.8) [{}] (5.1,-0.2) [{}]};
		\end{axis}
	\end{tikzpicture}
	\caption{Overhead sources of {\ToolName} (O: Observation, E: Enforcement)}
	\label{fig:rq4_overhead_sources}
\end{figure}

In enforcement mode, the average overhead of {\ToolName} across all step types is $7.9\%$, with JavaScript actions incurring the highest overhead at $8.5\%$.
To understand overhead sources, we decompose the total enforcement overhead into five components:
(1) Control-channel communication, which is the time taken to activate and deactivate the SIT;
(2) TLS interception, which captures the time required by the broker to complete the TLS handshake and establish the intercepted connection;
(3) Request verification, which is the time taken to authenticate the sender and verify its active SIT;
(4) Policy writing, which is the time taken to record the observed policy at the broker; and
(5) Remaining overhead, which accounts for the step's own execution and the additional proxy-relay overhead.
As shown in Figure~\ref{fig:rq4_overhead_sources}, the proxy-relay category accounts for $90.1\%$ of the enforcement overhead. 
TLS interception, control-channel communication, and request verification contribute $5.4\%$, $2.8\%$, and $1.6\%$ overhead, respectively.
The proxy relay captures the additional proxy hop, connection management, and the forwarding of requests and responses~\cite{298565,zhu2023meshinsight}.

Observation mode additionally stores each observed request and its inferred permissions in the policy file. 
These I/O operations increase the average overhead to $16.5\%$, compared with $7.9\%$ in enforcement mode. 
Policy writes account for $64.3\%$ of the observation mode overhead, while the proxy relay accounts for $32.2\%$.
The effect is most pronounced for composite actions, for which the observation mode overhead reaches $51.5\%$. 
Policy writes account for $93.0\%$ of this overhead, as network-intensive composite actions generate numerous observations that the broker must record. 
Consequently, the cost of these writes accumulates across the multiple actions within each composite action.
\section{Limitations}\label{sec:limitations}

\tightpar{Coverage of dynamic analysis}
{\ToolName}'s observation mode records the permissions only relevant to the execution paths taken during the runs, so an observed policy may miss a permission that a step needs on a path that was never taken. 
In our evaluation, we \emph{reduce} this risk by manually identifying and triggering all events and execution conditions of each workflow (Section~\ref{Sub:RQ2}). 
Static analysis tools can also provide an initial policy for the developer to review. 
We stress that this affects the \emph{inference} of policies and not their \emph{enforcement} as every request to the GitHub API still passes through the request verifier, and a missing permission results in a denied and logged request rather than a silent exposure of the repository to attacks. 
The developer can then see that denial in the workflow logs and update the policy.

\tightpar{Policy review}
Like any permission system, {\ToolName} assumes that the developer reviews the policy before enforcing it. 
An observed policy is a \emph{proposal}, rather than a proof of least privilege. 
If an action misuses a permission during the observed runs, that permission still appears in the observed policy. 
We therefore recommend observing policies using trusted versions of the actions and encourage developers to review them, as they already do with job-level permissions.

\section{Related Work}\label{rw}
We compare {\ToolName} to related works and position our contributions in the broader context of CI/CD security.

\tightpar{GitHub permissions}%
Our work is inspired by GitHub Advisor~\cite{GithubPer}, which provides job-level permission monitoring via a network proxy. 
In contrast, {\ToolName} focuses on both monitoring and runtime enforcement of step-level permissions, and attributes each request to its step through a step identifier token, which job-level network proxies cannot do.
\textsc{Cosseter} \cite{cosseter} proposes static analysis to infer the permissions for GitHub actions. 
Step Security's Harden-Runner~\cite{HardenRunner} and BOLT~\cite{bolt} enforce network-level rules, while Step Security Knowledge Base~\cite{stepsecurityKnowledgeBase} uses static analysis to derive permissions for popular actions. 
While static analysis struggles with dynamic behavior and missing runtime context, often resulting in overprivileged permissions, we believe it complements {\ToolName}, providing initial permissions to be reviewed by developers.
Koishybayev~\etal~\cite{Characterizing} analyze the security of GitHub Actions and other CI/CD platforms, identifying prevalent vulnerabilities, including overprivileged repository access. 
Delicheh~\etal~\cite{Mitigating} and Li~\etal~\cite{CIServices} study the attack surface of GitHub Actions, including malicious actions and common workflow misconfigurations such as poor permission management and overprivileged jobs.  
While these works highlight the risks of overprivileged jobs, they do not provide an effective solution to permission misuse attack, as {\ToolName} does.

\tightpar{Security of GitHub Actions ecosystem}%
Argus~\cite{argus} uses static taint analysis to detect code injection attacks in GitHub Actions, while Zhang~\etal~\cite{Zhang} investigate the use of LLMs for generating secure GitHub workflows. 
Several studies, such as Benedetti~\etal~\cite{ASS}, Decan~\etal~\cite{DecanICSME}, Khatami~\etal~\cite{codesmell}, and Vassallo~\etal~\cite{CDLinter} focus on detecting workflow misconfigurations and security-related code smells by analyzing workflows for recurring flaw patterns. 
Gu~\etal~\cite{CInspector} introduce CInspector to analyze vulnerabilities across seven popular CI platforms. 
They report on the prevalence of overprivileged tokens which enables attack vectors such as privilege escalation and malicious code injection. 
Cimon~\cite{cinamon} is a dynamic runtime tool that uses eBPF to monitor and mitigate software supply chain attacks pertaining to secret exfiltration and unauthorized filesystem access. 
GitHub CodeQL~\cite{CodeInjection} and SLSA~\cite{SALSA} detect command-injection vulnerabilities and artifact-poisoning attacks in GitHub Actions, respectively. 
These works are complementary and further show the importance of fine-grained enforcement tools like {\ToolName}.

\tightpar{Least privilege enforcement}%
Runtime enforcement of least privilege permissions is not new and spans different domains. 
The Android permission system ~\cite{10.1145/2046707.2046779} is perhaps the most popular, while runtimes such as Node.js have recently introduced permission models~\cite{NodejsPermission} to restrict access to the file system.
Cornelissen~\etal~\cite{NodeShield25} and Ferreira~\etal~\cite{Ferreira} study runtime enforcement for Node.js applications to enforce capabilities for JavaScript modules. In JavaScript, Ahmadpanah~\etal~\cite{Mohammad}, realsm~\cite{realsm}, and Vasilakis~\etal~\cite{Vasilakis} enforce security through language-level proxies, whereas {\ToolName} withholds the token from untrusted code.

\section{Conclusion}
We presented {\ToolName}, a runtime enforcement tool securing GitHub Actions workflows.
{\ToolName} re-designs the GitHub Actions permissions model by enabling fine-grained permissions at step level, mitigating permission misuse attacks on repositories. 
We evaluated {\ToolName} on real-world workflows for compatibility, security, and performance, demonstrating attack feasibility and {\ToolName}'s effectiveness in preventing them with acceptable overhead.
Future work includes extending {\ToolName} to support access control for other secrets and integrating it with other CI/CD platforms.

\tightpar{Acknowledgements}
LLMs (ChatGPT) were used for editorial purposes in this manuscript.

\bibliographystyle{IEEEtran}
\bibliography{bibliography}

\iffull
  \appendices
  
\section{Ethical Considerations}
\tightpar{Dataset collection} During the evaluation, authors crawled 500 workflows across most-starred public repositories using GitHub Search APIs\footnote{\url{https://github.com/search.html}}. To assess the security of Granite, authors created a separate remote repository containing the source code of these repositories, under a private organization owned by the authors. 

\tightpar{Malicious actions} To create malicious versions of actions in Section~\ref{Sub:RQ3}, authors repeated the same process, pushing the source code of malicious actions to our private GitHub organization. We then restricted access to these action repositories so that they were accessible only within our selected repositories in the same organization, ensuring that no external user could access or invoke the modified actions. 
All original git remotes were removed from the main repositories, ensuring there was no link between the new remote and the original remote.

\tightpar{Disclosure of findings} We have reached out to GitHub to present Granite and our findings. They acknowledged the security benefits of Granite and step-level permissions. The current efforts are focused on implementing and deploying mitigations such as action immutability for direct and transitive dependencies. 

\begin{table}[t]%
	\centering
	\scriptsize
	\begin{threeparttable}
		\begin{tabular}{@{}l p{0.38\columnwidth}@{}}
			\textbf{Repository} & \textbf{\shortstack[l]{\{\#write\_permissions/\\\#total\_steps\}}} \\
			\midrule[1px]
			
			EbookFoundation/free-programming-books
			& \{0/10\} \\
			
			jwasham/coding-interview-university
			& \{1/7\} \\
			
			ollama/ollama
			& \{0/7\} \\
			
			airbnb/javascript
			& \{1/2\} \\
			
			f/awesome-chatgpt-prompts
			& \{1/6\} \\
			
			facebook/react-native
			& \{1/2\} \\
			
			yt-dlp/yt-dlp
			& \{1/4\} \\
			
			langchain-ai/langchain
			& \{1/4\} \\
			
			langgenius/dify
			& \{1/7\} \\
			
			kamranahmedse/developer-roadmap
			& \{1/5, 1/5\} \\
			
			electron/electron
			& \{0/3, 1/4\} \\
			
			mrdoob/three.js
			& \{1/11, 2/4\} \\

			tensorflow/tensorflow
			& \{1/3, 1/4, 1/3\} \\

			Significant-Gravitas/AutoGPT
			& \{1/4, 1/8, 1/4\} \\
			
			twbs/bootstrap
			& \{1/4, 1/6, 2/4\} \\
			
			n8n-io/n8n
			& \{0/5, 0/6, 1/8\} \\

			axios/axios
			& \{1/10, 1/12, 1/3, 1/12, 1/4\} \\
			
			nodejs/node
			& \{1/4, 1/8, 1/7, 1/6, 1/5, 1/3, 1/7, 1/5, 2/4, 1/5\} \\
			
			\bottomrule
		\end{tabular}
		\begin{tablenotes}
			\scriptsize
			\item \textbf{Note:} Each cell lists the per-job ratio of steps granted write
			access by {\ToolName} to those granted write access under GitHub's
			permission model, in job order.
		\end{tablenotes}
		\caption{Attack Surface Reduction: {\ToolName} vs.\ GitHub permission model}
		\label{tab:attack_surface_reduction}
	\end{threeparttable}
\end{table}

\section{RQ2}
\label{appx:rq2_comparison}
\subsection*{Job Selection}
\label{appx:RQ2selection}
We select the RQ2 jobs from the overprivileged jobs identified in RQ1 (Section~\ref{Sub:RQ1}) in three steps.

First, we retain only jobs that can be executed in our organization. We exclude jobs that require third-party secrets through the \texttt{secrets} context and jobs that require Windows or macOS runners, since {\ToolName} builds on a Linux-based runner.
Second, we greedily rank the remaining jobs to maximize the diversity of evaluated actions. After selecting each job, we recompute the ranking, prioritizing jobs that introduce the most previously unseen third-party actions, followed by previously unseen action types (JavaScript, composite, or Docker) and trigger events. Repository stars are used only to break ties.
Repository local actions are executed but do not contribute to the ranking.

\subsection*{{\ToolName} vs GitHub Advisor}
We compare the job-level permissions reported by GitHub GHAdvisor with the union of the step-level permissions {\ToolName} observed for the same job, on the 30 jobs both tools executed (Figure~\ref{fig:rq2_advisor}).

\tightpar{Mismatches}
Three jobs differ, and in all of them GHAdvisor misses a permission that the job used. A policy derived from GHAdvisor would therefore break these jobs under enforcement.
In two jobs (\emph{zhayujie/CowAgent} and \emph{firecrawl/firecrawl}), a step pushes a container image to the GitHub container registry (\texttt{ghcr.io}) and requires \texttt{packages:write}. The push exchanges the job token for a registry-specific token and uploads the image with the latter. GHAdvisor records only requests to \texttt{github.com} and \texttt{api.github.com} that carry the job token, so it misses the push. For \emph{CowAgent} it reports only \texttt{packages:read}, which stems from another step that lists package versions through the REST API. {\ToolName} mediates the registry's token exchange and records the requested scope.
In the third job (\emph{ReactiveX/RxJava}), the Docker action \texttt{EnricoMi/publish-unit-test-result-action} creates a check run and reads the pull request hence needs (\texttt{checks:write}, \texttt{pull-requests:read}). GHAdvisor redirects only the traffic of processes on the runner host, using \texttt{iptables} rules on the \texttt{OUTPUT} chain, whereas the traffic of a container is forwarded through Docker's bridge and bypasses this redirect. GHAdvisor therefore recorded no permission for the action, although the check run was created. {\ToolName} passes its proxy configuration into the container and observes the action's requests.

\subsection*{{\ToolName} vs \textsc{Cosseter}}
\label{appx:rq2_cosseter_omissions}
\tightpar{Omission cases}
\textsc{Cosseter} omits a permission that {\ToolName} observed in three (job, action) pairs, each involving a different action (Figure~\ref{fig:rq2_static_comparison}).
In all three cases, the request that requires the permission is not visible as a GitHub API call in the action's JavaScript source.

\emph{\textbf{Child processes:}}
\emph{doocs/advanced-java} repository uses an action called \texttt{ad-m/github-push-action} that is a JavaScript action whose entry point only spawns a Bash script as a child process and the script commits the changes and runs \texttt{git push}. There is a similar case in the \emph{tw93/Mole} repository that uses \texttt{stefanzweifel/git-auto-commit-action} action. Both actions' push reaches GitHub through Git's HTTP protocol rather than the REST API, and requires \texttt{contents:write}.
\textsc{Cosseter} analyzes the JavaScript code, sees a command execution instead of a GitHub API call, and reports no permission for either action.
{\ToolName} observes the push regardless of the program that issues it and records \texttt{contents:write}.

\emph{\textbf{Indirect API call:}}
\texttt{dorny/paths-filter} from \emph{juanfont/headscale} lists the files changed by a pull request through the REST API, which requires \texttt{pull-requests:read}.
Instead of calling \texttt{pulls.listFiles} directly, the action builds the request options with \texttt{client.endpoint.merge(\dots)} and passes them to Octokit's helper \texttt{client.paginate.iterator}, which issues the requests.
\textsc{Cosseter} does not resolve this indirect call and reports no permission, which is also the result listed for this action in \textsc{Cosseter}'s own dataset.
The call is executed only for pull-request events when a token is provided, which is the case in this job. For other events, the action uses \texttt{git diff} and needs no permission.

\tightpar{Addition cases}
Table~\ref{tab:cosseter_additions} lists the ten (job, action) pairs in which \textsc{Cosseter} reports a permission that {\ToolName} did not observe.

\begin{table}[t]
	\centering
	\scriptsize
	\caption{Permission additions of \textsc{Cosseter}}
	\label{tab:cosseter_additions}
	\setlength{\tabcolsep}{3pt}
	\resizebox{\columnwidth}{!}{%
	\begin{tabular}{@{}l c >{\raggedright\arraybackslash}p{2.35cm} l@{}}
		\toprule
		\textbf{Action} & \textbf{Pairs} & \textbf{Added permission} & \textbf{Reason} \\
		\midrule
		\texttt{actions/upload-artifact} & 6 & \texttt{actions:write} & never needed \\
		\texttt{peter-evans/create-pull-request} & 1 & \texttt{contents:write}, \texttt{issues:write}, \texttt{pull-requests:write} & path not taken \\
		\texttt{styfle/cancel-workflow-action} & 1 & \texttt{actions:write} & path not taken \\
		\texttt{actions/setup-go} & 1 & \texttt{contents:read} & path not taken \\
		\texttt{actions/labeler} & 1 & \texttt{contents:read} & path not taken \\
		\bottomrule
	\end{tabular}}
\end{table}

\begin{figure*}
	\begin{tikzpicture}[scale=0.8, transform shape]
		\node[draw=blue!50, fill=blue!7, thick, rounded corners, minimum width=100pt] (workflowTriggers) at (-1.0,0) 
		{\scriptsize\begin{tabular}{@{\hspace{5pt}}>{\raggedright\arraybackslash}p{90pt}@{\hspace{5pt}}}
			\centering\textbf{WORKFLOW TRIGGERS} \tabularnewline
			\hline
			pull\_request \\
			push \\
			release \\
			workflow\_dispatch
		\end{tabular}};

		\node[draw=green!50, fill=green!7, thick, rounded corners, minimum width=100pt] (job1) at (3.1,0) 
		{\scriptsize\begin{tabular}{@{\hspace{5pt}}>{\raggedright\arraybackslash}p{90pt}@{\hspace{5pt}}}
			\centering\textbf{JOB 1: SETUP} \tabularnewline
			\hline
			1. actions/checkout*\\
			2. Check run condition\\
			3. Report pending status
		\end{tabular}};
		
		\node[draw=orange!50, fill=orange!7, thick, rounded corners, minimum width=150pt] (job2) at (9,1.5) 
		{\scriptsize\begin{tabular}{@{\hspace{5pt}}>{\raggedright\arraybackslash}p{130pt}@{\hspace{5pt}}}
				\centering\textbf{JOB 2: SECURITY \& QUALITY} \tabularnewline
				\hline
			1. actions/checkout*\\
			2. github/codeql-action/init*\\
			3. github/codeql-action/autobuild*\\
			4. github/codeql-action/analyze*\\
			5. lycheeverse/lychee-action*
		\end{tabular}};
		\node[draw=orange!50, fill=orange!7, thick, rounded corners, minimum width=150pt] (job4) at (9,3.8) 
		{\scriptsize\begin{tabular}{@{\hspace{5pt}}>{\raggedright\arraybackslash}p{130pt}@{\hspace{5pt}}}
			\centering\textbf{JOB 4: TEST \& COVERAGE} \tabularnewline
			\hline
			1. actions/checkout*\\
			2. actions/setup-python*\\
			3. Install deps\\
			4. Run tests\\
			5. coverallsapp/github-action*
		\end{tabular}};
		\node[draw=orange!50, fill=orange!7, thick, rounded corners, minimum width=150pt] (job5) at (9,6.1) 
		{\scriptsize\begin{tabular}{@{\hspace{5pt}}>{\raggedright\arraybackslash}p{130pt}@{\hspace{5pt}}}
			\centering\textbf{JOB 5: SBOM GENERATION} \tabularnewline
			\hline
			1. actions/checkout*\\
			2. Create repo list file\\
			\hangindent=1.15em\hangafter=1 3.~Brend-Smits/\allowbreak github-sbom-generator-action*\\
			4. actions/upload-artifact*\\
			5. peter-evans/create-or-update-comment*
		\end{tabular}};
		
		\node[draw=orange!50, fill=orange!7, thick, rounded corners, minimum width=150pt] (job3) at (9,-1.3) 
		{\scriptsize\begin{tabular}{@{\hspace{5pt}}>{\raggedright\arraybackslash}p{130pt}@{\hspace{5pt}}}
			\centering\textbf{JOB 3: AUDIT CHANGES} \tabularnewline
			\hline
			1. actions/checkout*\\
			2. Create config for auditor-action\\
			3. GrantBirki/auditor-action*\\
			4. codelytv/pr-size-labeler*\\
			\text{[PR only]}
		\end{tabular}};
		\node[draw=orange!50, fill=orange!7, thick, rounded corners, minimum width=150pt] (job6) at (9,-3.4) 
		{\scriptsize\begin{tabular}{@{\hspace{5pt}}>{\raggedright\arraybackslash}p{130pt}@{\hspace{5pt}}}
			\centering\textbf{JOB 6: CHANGELOG} \tabularnewline
			\hline
			1. actions/checkout*\\
			2. Create config\\
			3. Bibo-Joshi/chango*\\
			\text{[PR only]}
		\end{tabular}};
		\node[draw=orange!50, fill=orange!7, thick, rounded corners, minimum width=150pt] (job7) at (9,-5.95) 
		{\scriptsize\begin{tabular}{@{\hspace{5pt}}>{\raggedright\arraybackslash}p{130pt}@{\hspace{5pt}}}
			\centering\textbf{JOB 7: RELEASE NOTES} \tabularnewline
			\hline
			1. actions/checkout*\\
			2. Prepare release changelog\\
			3. ffurrer2/extract-release-notes*\\
			4. softprops/action-gh-release*\\
			5. Build distributions\\
			6. Attach to release\\
			\text{[Release only]}
		\end{tabular}};
		\node[draw=orange!50, fill=orange!7, thick, rounded corners, minimum width=150pt] (jobLint) at (9,-8.4) 
		{\scriptsize\begin{tabular}{@{\hspace{5pt}}>{\raggedright\arraybackslash}p{130pt}@{\hspace{5pt}}}
			\centering\textbf{JOB 8: LINT} \tabularnewline
			\hline
			1. actions/checkout*\\
			2. Lint with ruff\\
			3. Report lint status
		\end{tabular}};
		\node[draw=orange!50, fill=orange!7, thick, rounded corners, minimum width=150pt] (jobWfLint) at (9,-10.4) 
		{\scriptsize\begin{tabular}{@{\hspace{5pt}}>{\raggedright\arraybackslash}p{130pt}@{\hspace{5pt}}}
			\centering\textbf{JOB 9: WORKFLOW LINT} \tabularnewline
			\hline
			1. actions/checkout*\\
			2. docker://rhysd/actionlint*
		\end{tabular}};

		\begin{scope}[on background layer]
			\node[thick, rounded corners=5, inner sep=7, fit=(job5) (job4) (job2)] (box1) {};
			\node[thick, rounded corners=5, inner sep=7, fit=(job3) (job6)] (box2) {};
			\node[thick, rounded corners=5, inner sep=7, fit=(jobLint) (jobWfLint)] (box3) {};
			
			\draw[thick] (workflowTriggers.east) -- (job1.west);
			
			\draw[thick, rounded corners=10pt] 
				let
					\p1 = (job1.east),
					\p2 = (box1.west)
				in
					(\p1) --
					($(\x1,\y1)+(0.3,0)$) --
					($(\x1,\y2)+(0.3,0)$) --
					(\p2);
					
			\draw[thick, rounded corners=10pt] 
				let
					\p1 = (job1.east),
					\p2 = (box2.west)
				in
					(\p1) --
					($(\x1,\y1)+(0.3,0)$) --
					($(\x1,\y2)+(0.3,0)$) --
					(\p2);
			
			\draw[thick, rounded corners=10pt] 
			let
				\p1 = (job1.east),
				\p2 = (job7.west)
			in
				(\p1) --
				($(\x1,\y1)+(0.3,0)$) --
				($(\x1,\y2)+(0.3,0)$) --
				(\p2);
			
			\draw[thick, rounded corners=10pt] 
			let
				\p1 = (job1.east),
				\p2 = (box3.west)
			in
				(\p1) --
				($(\x1,\y1)+(0.3,0)$) --
				($(\x1,\y2)+(0.3,0)$) --
				(\p2);
		\end{scope}
	
		\node[draw=yellow!50, fill=yellow!7, thick, rounded corners, minimum width=150pt] (job8) at ($(box1.east)+(3.5,0)$) 
		{\scriptsize\begin{tabular}{@{\hspace{5pt}}>{\raggedright\arraybackslash}p{130pt}@{\hspace{5pt}}}
			\centering\textbf{JOB 10: AUTO-UPDATE PR} \tabularnewline
			\hline
			1. actions/checkout*\\
			2. Check dependency updates\\
			3. gr2m/create-or-update-pull-request-action*\\
			\text{[Workflow-dispatch only]}
		\end{tabular}};
	
		\node[draw=yellow!50, fill=yellow!7, thick, rounded corners, minimum width=150pt] (job9) at ($(box2.east)+(3.5,2.3)$) 
		{\scriptsize\begin{tabular}{@{\hspace{5pt}}>{\raggedright\arraybackslash}p{130pt}@{\hspace{5pt}}}
			\centering\textbf{JOB 11: WORKFLOW SUMMARY} \tabularnewline
			\hline
			1. Collect results \\
			2. peter-evans/create-or-update-comment* \\
			\text{[PR only]}
		\end{tabular}};
	
		\node[draw=yellow!50, fill=yellow!7, thick, rounded corners, minimum width=150pt] (jobStatus) at ($(box3.east)+(3.5,0)$) 
		{\scriptsize\begin{tabular}{@{\hspace{5pt}}>{\raggedright\arraybackslash}p{130pt}@{\hspace{5pt}}}
			\centering\textbf{JOB 12: REPORT STATUS} \tabularnewline
			\hline
			1. Report final status
		\end{tabular}};
	
		\begin{scope}[on background layer]
			\draw[thick] (box1.east) -- (job8.west);
			\draw[thick, rounded corners=10pt] 
				let
					\p1 = (box2.east),
					\p2 = (job9.west)
				in
					(\p1) --
					($(\x1,\y1)+(0.3,0)$) --
					($(\x1,\y2)+(0.3,0)$) --
					(\p2);
			\draw[thick, rounded corners=10pt] 
				let
					\p1 = (box3.east),
					\p2 = (job9.west)
				in
					(\p1) --
					($(\x1,\y1)+(0.3,0)$) --
					($(\x1,\y2)+(0.3,0)$) --
					(\p2);
			\draw[thick] (box3.east) -- (jobStatus.west);
			\draw[thick, rounded corners=10pt] 
				let
					\p1 = (box1.east),
					\p2 = (jobStatus.west)
				in
					(\p1) --
					($(\x1,\y1)+(0.7,0)$) --
					($(\x1,\y2)+(0.7,0)$) --
					(\p2);
			\draw[thick, rounded corners=10pt] 
				let
					\p1 = (box1.east),
					\p2 = (job9.west)
				in
					(\p1) --
					($(\x1,\y1)+(0.3,0)$) --
					($(\x1,\y2)+(0.3,0)$) --
					(\p2);
		\end{scope}

		\node[fill=black, draw=blue!7, thick, circle, inner sep=0, minimum size=5pt] at (workflowTriggers.east) {};
		
		\node[fill=black, draw=green!7, thick, circle, inner sep=0, minimum size=5pt] at (job1.east) {};
		\node[fill=black, draw=green!7, thick, circle, inner sep=0, minimum size=5pt] at (job1.west) {};
		
		\node[fill=black, draw=gray!10, thick, circle, inner sep=0, minimum size=5pt] at (box1.east) {};
		\node[fill=black, draw=gray!10, thick, circle, inner sep=0, minimum size=5pt] at (box1.west) {};
		\node[fill=black, draw=gray!10, thick, circle, inner sep=0, minimum size=5pt] at (box2.east) {};
		\node[fill=black, draw=gray!10, thick, circle, inner sep=0, minimum size=5pt] at (box2.west) {};
		\node[fill=black, draw=orange!7, thick, circle, inner sep=0, minimum size=5pt] at (job7.west) {};
		
		\node[fill=black, draw=yellow!7, thick, circle, inner sep=0, minimum size=5pt] at (job8.west) {};
		\node[fill=black, draw=yellow!7, thick, circle, inner sep=0, minimum size=5pt] at (job9.west) {};
		\node[fill=black, draw=yellow!7, thick, circle, inner sep=0, minimum size=5pt] at (jobStatus.west) {};
		\node[fill=black, draw=gray!10, thick, circle, inner sep=0, minimum size=5pt] at (box3.east) {};
		\node[fill=black, draw=gray!10, thick, circle, inner sep=0, minimum size=5pt] at (box3.west) {};
		
		\begin{scope}[on background layer]
			\node[draw=blue!50, very thick, circle, inner sep=0, minimum size=6pt] at (workflowTriggers.east) {};
			
			\node[draw=green!50, very thick, circle, inner sep=0, minimum size=6pt] at (job1.east) {};
			\node[draw=green!50, very thick, circle, inner sep=0, minimum size=6pt] at (job1.west) {};
			
			\node[draw=gray!50, very thick, circle, inner sep=0, minimum size=6pt] at (box1.east) {};
			\node[draw=gray!50, very thick, circle, inner sep=0, minimum size=6pt] at (box1.west) {};
			\node[draw=gray!50, very thick, circle, inner sep=0, minimum size=6pt] at (box2.east) {};
			\node[draw=gray!50, very thick, circle, inner sep=0, minimum size=6pt] at (box2.west) {};
			\node[draw=orange!50, very thick, circle, inner sep=0, minimum size=6pt] at (job7.west) {};
			
			\node[draw=yellow!50, very thick, circle, inner sep=0, minimum size=6pt] at (job8.west) {};
			\node[draw=yellow!50, very thick, circle, inner sep=0, minimum size=6pt] at (job9.west) {};
			\node[draw=yellow!50, very thick, circle, inner sep=0, minimum size=6pt] at (jobStatus.west) {};
			\node[draw=gray!50, very thick, circle, inner sep=0, minimum size=6pt] at (box3.east) {};
			\node[draw=gray!50, very thick, circle, inner sep=0, minimum size=6pt] at (box3.west) {};
			
			\node[fill=gray!10, draw=gray!50, thick, rounded corners=5, inner sep=7, fit=(job5) (job4) (job2)] {};
			\node[fill=gray!10, draw=gray!50, thick, rounded corners=5, inner sep=7, fit=(job3) (job6)] {};
			\node[fill=gray!10, draw=gray!50, thick, rounded corners=5, inner sep=7, fit=(jobLint) (jobWfLint)] {};
		\end{scope}
		
	\end{tikzpicture}
	\caption{The structure of workflow for performance evaluation.}
	\label{fig:performance_workflow}
\end{figure*}
\section{RQ3}

\begin{figure}[t]
	\centering
	\pgfplotsset{
		compat=1.18,
		rq4bar/.style={draw=black, line width=0.3pt},
		rq4legend/.style={legend image code/.code={\draw[##1] (0cm,-0.08cm) rectangle (0.32cm,0.08cm);}},
	}
	\begin{tikzpicture}[trim axis left, trim axis right]
		\begin{axis}[
			xbar, scale only axis,
			width=4.9cm, height=4.6cm,
			bar width=6.5pt,
			xmin=0, xmax=62, ymin=-0.6, ymax=4.6,
			xtick={0,10,20,30,40,50,60},
			ytick={0,1,2,3,4},
			yticklabels={Docker action, Composite, JavaScript, Run, \textbf{All steps}},
			yticklabel style={xshift=-4pt, text width=1.4cm, align=right, font=\scriptsize},
			ytick style={draw=none},
			tick label style={font=\footnotesize},
			xlabel={Overhead relative to unmodified runner (\%)},
			xlabel style={font=\footnotesize},
			xmajorgrids, grid style={gray!25},
			clip=false,
			title={\textbf{Overhead per step type}},
			title style={font=\small},
		]
			\addplot[rq4bar, fill=gray!60, bar shift=3.6pt] coordinates {(16.516,4) (7.207,3) (11.349,2) (51.484,1) (4.873,0)};
			\addplot[rq4bar, fill=gray!20, bar shift=-3.6pt] coordinates {(7.903,4) (6.618,3) (8.543,2) (6.387,1) (1.868,0)};
			\node[anchor=west, font=\fontsize{6}{7}\selectfont, inner sep=1pt] at (axis cs:16.516,4.2) {16.5\%};
			\node[anchor=west, font=\fontsize{6}{7}\selectfont, inner sep=1pt] at (axis cs:7.903,3.8) {7.9\%};
			\node[anchor=west, font=\fontsize{6}{7}\selectfont, inner sep=1pt] at (axis cs:7.207,3.2) {7.2\%};
			\node[anchor=west, font=\fontsize{6}{7}\selectfont, inner sep=1pt] at (axis cs:6.618,2.8) {6.6\%};
			\node[anchor=west, font=\fontsize{6}{7}\selectfont, inner sep=1pt] at (axis cs:11.349,2.2) {11.3\%};
			\node[anchor=west, font=\fontsize{6}{7}\selectfont, inner sep=1pt] at (axis cs:8.543,1.8) {8.5\%};
			\node[anchor=west, font=\fontsize{6}{7}\selectfont, inner sep=1pt] at (axis cs:51.484,1.2) {51.5\%};
			\node[anchor=west, font=\fontsize{6}{7}\selectfont, inner sep=1pt] at (axis cs:6.387,0.8) {6.4\%};
			\node[anchor=west, font=\fontsize{6}{7}\selectfont, inner sep=1pt] at (axis cs:4.873,0.2) {4.9\%};
			\node[anchor=west, font=\fontsize{6}{7}\selectfont, inner sep=1pt] at (axis cs:1.868,-0.2) {1.9\%};
			\node[anchor=east, font=\tiny, text=black!65, inner sep=1pt] at (axis cs:0,4.2) {O};
			\node[anchor=east, font=\tiny, text=black!65, inner sep=1pt] at (axis cs:0,3.8) {E};
			\node[anchor=east, font=\tiny, text=black!65, inner sep=1pt] at (axis cs:0,3.2) {O};
			\node[anchor=east, font=\tiny, text=black!65, inner sep=1pt] at (axis cs:0,2.8) {E};
			\node[anchor=east, font=\tiny, text=black!65, inner sep=1pt] at (axis cs:0,2.2) {O};
			\node[anchor=east, font=\tiny, text=black!65, inner sep=1pt] at (axis cs:0,1.8) {E};
			\node[anchor=east, font=\tiny, text=black!65, inner sep=1pt] at (axis cs:0,1.2) {O};
			\node[anchor=east, font=\tiny, text=black!65, inner sep=1pt] at (axis cs:0,0.8) {E};
			\node[anchor=east, font=\tiny, text=black!65, inner sep=1pt] at (axis cs:0,0.2) {O};
			\node[anchor=east, font=\tiny, text=black!65, inner sep=1pt] at (axis cs:0,-0.2) {E};
		\end{axis}
	\end{tikzpicture}
	\caption{Overhead of {\ToolName} per step type relative to the unmodified runner, in observation mode (O) and enforcement mode (E)}
	\label{fig:rq4_overhead_steptype}
\end{figure}

\subsection*{Attack Implementation}
\label{appx:attack_implementation}
Each payload in Table~\ref{tab:security_analysis} first performs its attack with the credential the step receives, and then applies one of the techniques. The payloads of JavaScript actions use Node.js modules, and those of composite and Docker actions use shell commands and \texttt{curl}.

\tightpar{T1: Reading the job token}
The payload searches every place where the unmodified runner exposes \texttt{GITHUB\_TOKEN} to a step. It declares an optional input with the default value \texttt{\$\{\{ github.token \}\}}, reads \texttt{\$\{\{ secrets.GITHUB\_TOKEN \}\}} in the workflow, and dumps the step's environment variables and the files the runner writes for the step. Under {\ToolName}, every one of these places holds the step's SIT, because the runner removes the token from the \texttt{github} and \texttt{secrets} contexts when the job arrives. The payload then uses the value it found to call the API endpoint of its attack.
We apply this technique in the \emph{environment variable exfiltration} and \emph{dependency substitution} attacks (AutoGPT).

\tightpar{T2: Forging the control channel}
The payload tries to obtain a new SIT with a broader policy by acting as the runner's worker. It looks for the control channel among its inherited file descriptors, connects to the credential broker's control socket, and sends a job-registration and a SIT-issuance request. It also tries to read the worker's memory through \texttt{/proc/<pid>/mem} to recover the session. The worker does not pass the channel to step processes, the credential broker rejects a second session while the worker's session is active, and the worker's memory is not dumpable, so reading it fails.
We apply this technique in the \emph{LLM API key exfiltration} attack (dify).

\tightpar{T3: Reusing a stored SIT}
The payload reuses a SIT that an earlier step left on disk, such as the credential \texttt{actions/checkout} persists in the repository's Git configuration. It reads the stored value and sends its requests with it instead of its own SIT. The credential broker attributes each request to the step that sends it, identified by the step's proxy credential, and checks the request against that step's policy. A stored SIT therefore grants no permission beyond the policy of the step that uses it.
We apply this technique in the \emph{backdoor injection} (axios) and \emph{attach fake security checks} (bootstrap) attacks.

\tightpar{T4: Bypassing the proxy}
The payload sends its requests directly to \texttt{api.github.com}, without passing through the {\ToolName} controller. It clears the proxy environment variables and opens a TLS connection to GitHub itself (with \texttt{curl --noproxy '*'} in composite and Docker actions, and a raw \texttt{tls} socket in JavaScript actions, which the routing shim does not cover). These requests carry only the step's SIT, which GitHub does not accept as a credential.
We apply this technique in the \emph{unit test bypass} (axios) and \emph{dismiss security alerts} (Node.js) attacks.

The remaining three attacks (\emph{prompts modification}, \emph{approve and merge pull-request}, and \emph{phishing via issue comments}) use the credential they receive as is, without attempting to bypass {\ToolName}. 

\subsection*{Attack Surface Reduction}
\label{appx:attack_reduction}
Table \ref{tab:attack_surface_reduction} shows the detailed results of attack surface reduction by {\ToolName} compared to current GitHub job-level permission model which is discussed in Section \ref{Sub:RQ3}.
\section{Performance Overhead}
\label{appx:performance}
Figure \ref{fig:performance_workflow} shows the structure of the workflows used in the performance evaluation in Section \ref{sec:performance_overhead}. Figure~\ref{fig:rq4_overhead_steptype} illustrates the overhead of {\ToolName} per step type, and Figure~\ref{fig:rq4_overhead_sources} shows the sources of the overhead.

\fi

\end{document}